\documentclass[aps,prd,eqsecnum,preprintnumbers,showpacs]{revtex4}
\usepackage{bm}
\usepackage{amssymb}
\usepackage{graphicx}
\newcommand{\hatE}{\hat{E}}

\newcommand{\dphi}{\dot{\phi}}
\newcommand{\ddphi}{\ddot{\phi}}

\begin{document}
\hspace{-10mm}
\leftline{
\includegraphics{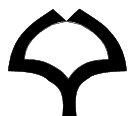}
}
\vspace*{-10.0mm} 
\thispagestyle{empty}
{\baselineskip-4pt
\font\yitp=cmmib10 scaled\magstep2
\font\elevenmib=cmmib10 scaled\magstep1  \skewchar\elevenmib='177
\leftline{\baselineskip20pt
\hspace{10mm} 
\vbox to0pt
   { {\yitp\hbox{Osaka \hspace{1.5mm} University} }
     {\large\sl\hbox{{Theoretical Astrophysics}} }\vss}}

\rightline{\large\baselineskip20pt\rm\vbox to20pt{
\baselineskip14pt
\hbox{OU-TAP-198}
\vspace{1mm}
}%
}
\vskip8mm
\begin{center}{\large\bf
 Geometry and cosmological perturbations in the bulk inflaton model}
\end{center}

\begin{center}
\large
Masato Minamitsuji$^{1,2}$,
Yoshiaki Himemoto$^3$
and
Misao Sasaki$^2$
\end{center}

\begin{center}
\em
$^1$Department of Earth and Space Science, Graduate School of Science,
\\
Osaka University, Toyonaka 560-0043, Japan
\end{center}
\begin{center}
\em
$^2$Yukawa Institute for Theoretical Physics,
\\
 Kyoto University, Kyoto 606-8502, Japan
\end{center}
\begin{center}
\em
$^3$Research Center for the Early Universe, School of Science,
\\
University of Tokyo, Tokyo 113-0033, Japan
\end{center}


\begin{abstract}
We consider a braneworld inflation model driven
by the dynamics of a scalar field living in the
5-dimensional bulk, the so-called ``bulk inflaton model'',
and investigate the geometry in the bulk and
large scale cosmological perturbations on the brane.
The bulk gravitational effects on the brane are
described by a projection of the 5-dimensional Weyl tensor,
which we denote by $E_{\mu\nu}$.
Focusing on a tachionic potential model, we take a perturbative approach 
in the anti-de Sitter (AdS$_5$) background
with a single de Sitter brane.
We first formulate the evolution equations for $E_{\mu\nu}$ in the bulk.
Next, applying them to the case of a spatially homogeneous brane,
we obtain two different integral expressions for $E_{\mu\nu}$.
One of them reduces to the expression
obtained previously when evaluated on the brane. The other
is a new expression that may be useful for analyzing the
bulk geometry. Then we consider superhorizon scale
cosmological perturbations and evaluate the bulk effects
onto the brane. In the limit $H^2\ell^2\ll1$, where $H$ is the
Hubble parameter on the brane and $\ell$ is the bulk curvature
radius, we find that the effective theory on the brane
is identical to the 4-dimensional Einstein-scalar theory
with a simple rescaling
of the potential even under the presence of inhomogeneities.
 In particular,
it is found that the anticipated non-trivial bulk effect due to
the spatially anisotropic part of $E_{\mu\nu}$ may appear only at
$O(H^4\ell^4)$.
\end{abstract}


\date{\today}

\maketitle

\section{Introduction}

Recent progress in particle physics has lead to the idea that
our universe is a (mem)brane in a higher dimensional spacetime
and all the interactions but gravity are confined on the brane,
the so-called braneworld scenario~\cite{braneworld}.
In particular, the scenario proposed by
Randall and Sundrum~\cite{Randall:1999vf}, in which
a single positive tension brane is embedded in the 5-dimensional
anti-de Sitter space (AdS$_5$) with $Z_2$ symmetry (the RS2 scenario),
has attracted much attention as an alternative to the
standard 4-dimensional cosmology~\cite{bwcosreview}
because of its
attractive feature that the standard 4-dimensional gravity is recovered
on the brane in the low energy limit~\cite{Randall:1999vf,Garriga:1999yh}.
The inflationary cosmology is also widely accepted because of its success
in explaining cosmological observations \cite{KT}.
It is therefore natural to consider the inflationary cosmology
from the point of view of
the braneworld scenario~\cite{braneinf}.
Brane inflation in the quantum cosmological context
is also of great
interest~\cite{Garriga:1999bq,bwcreation}.

In this paper, we consider a model of braneworld inflation driven
by the dynamics of a scalar
field $\phi$ in the 5-dimensional
bulk~\cite{Himemoto:2000nd,bwqfluc,LanRod01,Himemoto:2001hu,bwreheat}
in the context of the RS2 scenario,
and investigate the cosmological perturbations on superhorizon
scales, i.e., in the long wavelength limit.
The existence of such a scalar field is supported from the unified
theoretical
point of view, as the reduction of a higher dimensional theory to
5 dimensions will give rise to scalar-tensor type effective theories.
For a relevant form of the potential in the bulk, the scalar field
dynamics can give rise to inflation on the brane~\cite{Himemoto:2000nd}.
A theoretically interesting aspect of this bulk inflaton model is
that the bulk is not inflating at all; inflation and the
subsequent hot Friedmann stage of the universe can be realized
solely by the geometrical dynamics of the bulk.
Furthermore, in the low energy limit $H\ell\ll1$ where $H$ is
the Hubble parameter of the brane and $\ell$ is the curvature radius
of the bulk, as far as the spatially homogeneous
dynamics is concerned, the dynamics on the brane has been
found to be indistinguishable form the conventional slow-roll
inflation~\cite{Himemoto:2001hu}.
It was also shown that the quantum fluctuations projected
on the brane~\cite{bwqfluc} as well as reheating
after inflation~\cite{bwreheat} also mimic the 4-dimensional standard
inflation, as long as we consider the low energy limit.

Thus the bulk inflaton model may be a viable alternative
to the conventional 4-dimensional inflationary scenario.
It is then of importance to clarify if the cosmological perturbations
produced in the bulk inflaton model also have the same desirable
features as those in the 4-dimensional theory in the low energy limit,
and if there exists any signature specific to the braneworld scenario
that can be observationally tested.
Very recently, cosmological perturbations in a bulk inflaton
model with an exponential potential and dilatonic coupling to
the brane tension have been investigated by
Koyama and Takahashi~\cite{KoyTak03}.
In this paper, we consider a tachionic bulk potential
with the potential maximum at $\phi=0$,
with no coupling to the brane tension, and
investigate the bulk geometry and
cosmological perturbations in this model.
We take the geometrical approach to the effective gravitational
equations on the brane developed by Shiromizu,
Maeda and Sasaki~\cite{Shiromizu:1999wj} and by
Maeda and Wands~\cite{Maeda:2000wr}.
In this formalism, the bulk gravitational effects on the brane are
described by the projected 5-dimensional Weyl tensor,
\begin{eqnarray}
E_{\mu\nu}
=\mathop{C}\limits^{(5)}\!\!{}_{\mu a\nu b}\,n^an^b\,,
\nonumber
\end{eqnarray}
where $n^a$ is the unit vector normal to the brane
(Latin indices run over 5 dimensions, while Greek indices
over 4 dimensions, with the choice of Gaussian normal
coordinates such that $n^a\partial_a=\partial_5$).

This paper is organized as follows.
In Sec.~II, we briefly review the bulk inflaton model,
based on the geometrical formalism~\cite{Shiromizu:1999wj,Maeda:2000wr}.
In Sec.~III, we derive the evolution equations for $E_{\mu\nu}$
in the bulk when there exists a non-trivial bulk energy momentum tensor,
and solve them in the case of a spatially homogeneous brane, under the
assumption that the amplitude of the scalar field $\phi$ is small.
In Sec.~IV, we consider the cosmological perturbations
on superhorizon scales on the brane by solving the perturbation
equations for $\phi$ and $E_{\mu\nu}$ in the bulk.
Since the tensor perturbations in this model are identical
to the vacuum AdS$_5$ bulk model at leading order, we
 focus on the scalar-type perturbations.
We find the standard 4-dimensional result is recovered
in the low energy limit $H^2\ell^2\ll1$, and possible signatures of
the braneworld appear only at $O(H^4\ell^4)$.
 In Sec.~V , we summarize our results and discuss
the implications. Some formulas used in the text
are summarized in Appendices~\ref{ProjTab} and \ref{Source}.
Properties of the Green function for $E_{\mu\nu}$
in the bulk are analyzed in Appendix~\ref{Greenfcn}.

\section{Review of the Bulk Inflaton Model}

First, we review the bulk inflaton model~\cite{Himemoto:2000nd}.
We choose the Gaussian normal coordinates,
\begin{eqnarray}
ds^2=(n_a n_b+q_{ab})dx^a dx^b
=dr^2+q_{\mu\nu}dx^{\mu}dx^{\nu}\,.
\label{bulkmetric}
\end{eqnarray}
We assume that the brane is
located at $r=r_0$.
By extremizing the action of the 5-dimensional Einstein-scalar
system with
a brane, the gravitational equations in the bulk take the form
\begin{eqnarray}
{}^{(5)}G_{ab}+\Lambda_{5} g_{ab}
=\kappa_{5}^2\Bigl(\,T_{ab}\bigl[\phi\bigr]+S_{ab} \delta(r-r_{0})\Bigr)
\label{5deq},
\end{eqnarray}
where
$\kappa_{5}^2$ and $\Lambda_{5}$ are
the 5-dimensional gravitational and cosmological constants,
respectively. $T_{ab} \bigl[\phi \bigr]$ is
 the energy-momentum tensor of the bulk scalar field
for which we assume the form,
\begin{eqnarray}
T_{ab}\bigl[\phi\bigr]= \partial_{a}\phi\partial_{b}\phi -g_{ab}
 \Bigl(\frac{1}{2} g^{cd} \partial_{c}\phi \partial_{d} \phi
+V [\phi] \Bigr),
\label{Tab}
\end{eqnarray}
and $S_{ab}$ is the energy-momentum tensor on the brane for which
we assume the vacuum form,
\begin{eqnarray}
S_{ab}=-\sigma q_{ab},
\label{Sab}
\end{eqnarray}
where $\sigma$ is the brane tension.
Following the spirit of the RS2 scenario, we
assume the tuning between $\Lambda_5$ and $\sigma$ as
\begin{eqnarray}
\Lambda_5=-\frac{\kappa_5^4\sigma^2}{6}=-\frac{6}{\ell_0^2}\,,
\end{eqnarray}
where $\ell_0$ is the AdS$_5$ curvature radius of the RS2 braneworld,
and assume the $Z_2$ symmetry with respect to the brane.
Furthermore, we assume that the potential takes
a tachionic form
\begin{eqnarray}
V(\phi)= V_{0}+\frac{1}{2}m^2 \phi^2\,; \quad V_0>0\,,
\quad m^2<0\,,
\label{bulkpot}
\end{eqnarray}
in the vicinity of $\phi=0$. We tacitly assume that
there exists a minimum of the potential somewhere at
$\phi=\phi_{\rm min}\neq0$ at which $V(\phi_{\rm min})=0$,
where the RS2 flat brane is recovered.

The field equation for $\phi$ in the bulk is
\begin{eqnarray}
\nabla^{a}\nabla_{a}\phi-\frac{d}{d\phi}V[\phi]=0,
\label{phieq}
\end{eqnarray}
where $\nabla_{a}$ is the 5-dimensional covariant derivative.
Here, for simplicity, we impose the Neumann boundary condition
\begin{eqnarray}
\partial_{r}\phi|_{r=r_0}=0\,,
\label{Neumann}
\end{eqnarray}
on the scalar field.
This implies that the scalar field does not
couple to the brane tension.

{}From Israel's junction condition on the brane,
\begin{eqnarray}
\Bigl[K_{\mu\nu} \Bigr] = -\kappa_{5}^2
\Bigl( S_{\mu\nu}-\frac{1}{3} q_{\mu\nu} S \Bigr)
= -\frac{1}{3}\kappa_{5}^2 \sigma q_{\mu\nu}\,,
\label{israel}
\end{eqnarray}
and the $Z_{2}$ symmetry,
we obtain the 4-dimensional effective gravitational equations
on the brane~\cite{Himemoto:2000nd,Shiromizu:1999wj}
\begin{eqnarray}
\mathop{G}\limits^{(4)}\!\!{}_{\mu\nu}
=\kappa_{5}^2 T^{\rm(b)}_{\mu\nu}-E_{\mu\nu},
\label{eff4deq}
\end{eqnarray}
 where
\begin{eqnarray}
&&T^{\rm (b)}_{\mu\nu}=\frac{2}{3}
\Bigl[T_{cd} q^{c}_a q^{d}_{b} + \Bigl(T_{cd}n^c n^d
-\frac{1}{4}T \Bigr)q_{ab}  \Bigr]\,,
\label{Tbmn}
\\
&&E_{\mu\nu}\equiv
\mathop{C}\limits^{(5)}\!\!{}_{abcd}
\,q^{a}_{\mu}\,q^{c}_{\nu}n^{b}n^{d}.
\label{Emndef}
\end{eqnarray}
${}^{(5)} C_{abcd}$ is five-dimensional Weyl tensor.

In general, the projected Weyl tensor term $E_{\mu\nu}$ cannot be
determined without solving the bulk dynamics.
However, for a spatially homogeneous and isotropic brane,
$E_{\mu\nu}$ on the brane may be evaluated without solving the bulk.
By using the 4-dimensional contracted Bianchi identities,
one finds~\cite{Himemoto:2000nd}
\begin{eqnarray}
E_{tt}=\frac{\kappa_{5}^2}{2a^4(t)}\int^{t}dt' a^4(t')
\Bigl(\partial_{r}^2\phi+\frac{\dot{a}}{a}\dot{\phi}\Bigr),
\label{Etthomo}
\end{eqnarray}
where $a(t)$ is the cosmic scale factor on the brane and
$t$ the cosmic proper time. The other components of $E_{\mu\nu}$
are determined by the isotropy of the brane and the traceless nature
of $E_{\mu\nu}$.

For the potential of the form (\ref{bulkpot}), we then
focus on the region $|m^2|\phi^2\ll V_0$ and
consider the perturbation with respect to the amplitude of $\phi$.
In the zeroth order, the background metric is determined
as~\cite{Garriga:1999bq}
\begin{eqnarray}
ds^2=b^2(z)(dz^2-dt^2+a(t)^2\delta_{ij} dx^{i}dx^{j}),
\label{bgmetric}
\\ \nonumber
b(z)=\frac{Hl}{\sinh\bigl[H(|z|+z_0)\bigr]},\quad a(t)
=\frac{e^{Ht}}{H}\,.
\end{eqnarray}
Here $dz=dr/b(r)$ and the brane is located at
$z=0$, $H\ell=\sinh[Hz_0]$, and
\begin{eqnarray}
\ell=\sqrt{-\frac{6}{\Lambda_5+\kappa_5^2V_0}}
\end{eqnarray}
is the effective ${\rm AdS_{5}}$ radius. Note that $\ell>\ell_0$.
The braneworld inflates with the Hubble rate
$H$ given by
\begin{eqnarray}
H^2 = \kappa_{5}^2\, \frac{V_{0}}{6}.
\label{bgHubble}
\end{eqnarray}

In the first order in $\phi$,
the evolution of the scalar field on the background
metric is determined.
Expressing $\phi$ as the sum over all possible modes,
$\phi=\sum_n u_n(z)\psi_n(t)$,
the late time behavior of $\phi$ on the brane
is dominated by the bound-state mode for which the
effective 4-dimensional mass-squared for $\phi(t)$ is
given by~\cite{Himemoto:2000nd}
\begin{eqnarray}
M_{\rm{eff}}^2 \approx \frac{m^2}{2}\,,
\label{Meff}
\end{eqnarray}
for $H^2l^2 \ll 1$ and $|m^2|/H^2 \ll 1$.
In this case, one has
\begin{eqnarray}
\partial^2_z\phi|_b={m^2\over2}\phi=-\ddot\phi-3H\dot\phi\,,
\label{d2zphi}
\end{eqnarray}
on the brane.

The effective equations (\ref{eff4deq}) at second order
is then derived as follows.
Using Eq.~(\ref{d2zphi}), $E_{tt}$ is evaluated as
\begin{eqnarray}
E_{tt} =-\frac{\kappa_{5}^2}{2 a^4}\int^{t}dt\, a^4 \dot{\phi}
(\ddot{\phi}+2H\dot{\phi})
 =-\frac{\kappa_{5}^2}{4}\dot{\phi}^2+{C\over a^4}\,,
\label{Ettlate}
\end{eqnarray}
where $C$ is an integration constant which depends on the initial condition.
The term $C/a^4$ is called dark radiation term. In the present
case, since we are interested in the late time behavior
at the inflationary
stage, the dark radiation term can be safely neglected.
Then, the effective Friedmann equation becomes
\begin{eqnarray}
3\Bigl(\frac{\dot{a}}{a}\Bigr)^2
= \kappa_{4}^2\,\rho^{\rm (b)}-E_{tt}
\equiv\kappa_4^2\,\rho_{\rm eff}\,,
\label{effFeq}
\end{eqnarray}
where
\begin{eqnarray}
\rho^{\rm (b)}=\ell_0\left({1\over4}\dot\phi^2+{1\over2}V(\phi)\right);
\quad \kappa_4^2=\frac{\kappa_5^4\sigma}{6}=\frac{\kappa_5^2}{\ell_0}\,,
\end{eqnarray}
and
\begin{eqnarray}
\rho_{\rm{eff}}=\ell_0
 \left(\frac{1}{2}\dot{\phi}^2 +\frac{1}{2}V(\phi)\right).
\label{rhoeff}
\end{eqnarray}
Therefore, by introducing the effective 4-dimensional field
$\varphi$ as
\begin{eqnarray}
\varphi = \sqrt{l_{0}}\phi\,,
\end{eqnarray}
the system is equivalent to the 4-dimensional
Einstein-scalar system with
the potential,
\begin{eqnarray}
U(\varphi)= \frac{\ell_{0}}{2} V(\varphi/\sqrt{\ell_{0}})\,.
\end{eqnarray}
Thus the effective field $\varphi$ behaves as a
conventional inflaton on the brane, and
the conventional slow-roll inflation is realized on the brane
in the low energy approximation
$H^2\ell^2\ll1$~\cite{Himemoto:2000nd,Himemoto:2001hu}.

\section{Evolution Equations in the Bulk}

When the bulk scalar field is spatially inhomogeneous,
it is necessary to solve the bulk geometry to evaluate
$E_{\mu\nu}$ on the brane.
In this section, we derive the evolution equations
for $E_{\mu\nu}$. We then apply the resulting equations
to the spatially homogeneous and isotropic case to
see how the previous result reviewed in Sec.~II is
recovered from the bulk point of view.

\subsection{General equations}

First, we recapitulate the definitions of
the two projected Weyl tensors in the bulk,
\begin{eqnarray}
E_{ab}=\mathop{C}\limits^{(5)}\!\!
{}_{cedf}n^{c}n^{d}\,q^{e}_{a}\,q^{f}_{b}\,,\,\,
B_{abc}=q^{d}_{a}\,q^{e}_{b}\mathop{C}\limits^{(5)}\!\!
{}_{decf}n^{f}.
\label{EBdef}
\end{eqnarray}
By definition, $E_{ab}$ is symmetric with respect to the indices
($a,b$) whereas
$B_{abc}$ is anti-symmetric with respect to ($a,b$). Furthermore
$E_{ab}$ is traceless $E^{a}{}_{a}=0$.

We start from the 5-dimensional Bianchi identities
\begin{eqnarray}
\nabla_{[a}\mathop{R}\limits^{(5)}\!\!{}_{bc]de}=0,
\label{5bianchi}
\end{eqnarray}
where $\mathop{R}\limits^{(5)}\!\!{}_{abcd}$ is the 5-dimensional Riemann
tensor. We can derive the equations for $E_{\mu\nu}$
and $B_{abc}$ from Eq.~(\ref{5bianchi}) by using
the 5-dimensional Einstein equations~(\ref{5deq}).
We find
\begin{eqnarray}
{\pounds_{n}}E_{ab}
& =& D^{c}B_{c(ab)}
+K^{cd}\mathop{C}\limits^{(4)}\!\!{}_{acbd}
+4K^{c}_{(a}E_{b)c} -\frac{3}{2}KE_{ab}
-\frac{1}{2}q_{ab}K^{cd}E_{cd}
\nonumber\\
&&\qquad
+2\tilde{K}_{ac}\tilde{K}^{cd}\tilde{K}_{bd}
-\frac{7}{6}\tilde{K}_{ab}\tilde{K}^{cd}
\tilde{K}_{cd}
+\kappa_5^2\,P_{ab}\bigl[T_{ab}\bigr]
\label{LnE}
\end{eqnarray}
 where
\begin{eqnarray}
P_{ab}\bigl[T_{ab}\bigr]
&\equiv&
-\frac{1}{3}T_{de}n^{d}n^{e}q_{ab}K
+\frac{1}{3}TK_{ab}
-\frac{2}{3}K^{d}_{(a}q_{b)}^{e}
T_{de}
+\frac{1}{3}K^{de}T_{de}q_{ab}
\nonumber\\
&&\qquad
+\frac{2}{3}{\pounds_{n}}F_{ab}
-\frac{2}{3}D^{c}\Bigl(T_{de}n^{e}
q^{d}_{[c}q_{(a]b)} \Bigr)
-D_{(a}\Bigl[q^{d}_{b)}T_{de}n^{e}\Bigr]\,,
\label{Pdef}
\end{eqnarray}
with
\begin{eqnarray}
F_{ab}&=& T_{fg}q_{a}^{f}q_{b}^{g}
-\bigl(T_{fg}n^{f}n^{g} -\frac{1}{4}T \bigr)
q_{ab}-\frac{1}{2}T_{fg}q^{fg}q_{ab},
\nonumber\\
\tilde{K}_{ab}
& =& K_{ab}-\frac{1}{4}K q_{ab}\,,
\label{FtilKdef}
\end{eqnarray}
and
\begin{eqnarray}
{\pounds_{n}}B_{abc}=-2D_{[a}E_{b]c}
+K^{g}_{c}B_{abg}
-2B_{cg[a}K^{g}_{b]}
+\kappa_5^2\,Q_{abc}\bigl[T_{ab}\bigr],
\label{LnB}
\end{eqnarray}
 where
\begin{eqnarray}
Q_{abc}\bigl[T_{ab}\bigr]
& \equiv&
-\frac{2}{3}T_{de}n^{e}
\Bigl(q^{d}_{[a}K_{b]c}-K^{d}_{[a}q_{b]c}\Bigr)
+\frac{2}{3}{\pounds_{n}} \Bigl[T_{de}
n^{e}q^{d}_{[a}q_{b]c}  \Bigr]
\nonumber\\
&&\qquad
-\frac{2}{3}D_{[a}\Bigl[q_{b]c}\bigl(T_{de} n^{d}n^{e}
-\frac{T}{2}\bigr)\Bigr]
-\frac{2}{3}D_{[a}\Bigl[q_{b]}^{d}q_{c}^{e}T_{de}\Bigr] .
\label{Qdef}
\end{eqnarray}
They are first-order partial differential equations for
$E_{ab}$ and $B_{abc}$.

\subsection{Second order equations for $\bm{E_{\mu\nu}}$}

We assume the tachionic potential form~(\ref{bulkpot}), and take
the perturbation approach with respect to the amplitude of $\phi$.

First we expand the energy momentum tensor of the scalar field as
\begin{eqnarray}
T_{ab}[\phi]&=&T^{(0)}_{ab}[\phi]+T^{(2)}_{ab}[\phi]\,;
\nonumber\\
&&T^{(0)}_{ab}[\phi]=-V_{0}\,g_{ab}\,,
\nonumber\\
&&T^{(2)}_{ab}[\phi]=\partial_{a}\phi\partial_{b}\phi
-\frac{1}{2}g_{ab}\Bigl(g^{cd}\partial_{c}\phi\partial_{d}\phi
+m^2\phi^2\Bigr).
\label{Texpand}
\end{eqnarray}
So far, we have not specified the coordinate system. Now,
for definiteness, we choose the conformal
(Gaussian normal) coordinate~(\ref{bgmetric}).
The extrinsic curvature of each $z={\rm constant}$ hypersurface
is expanded as
\begin{eqnarray}
K_{\mu\nu}&=&K^{(0)}_{\mu\nu}+K^{(2)}_{\mu\nu}\,;
\nonumber\\
&&K^{(0)}_{\mu\nu}=\frac{1}{2b(z)}\partial_{z}q_{\mu\nu}
=\frac{b'(z)}{b^2(z)}q_{\mu\nu}\,,
\nonumber\\
&&K^{(2)}_{\mu\nu}=O(\phi^2),
\label{Kexpand}
\end{eqnarray}
corresponding to the amplitude of the scalar field.

Substituting the above expansions of $T_{ab}$ and $K_{\mu\nu}$
into Eqs.~(\ref{LnE}) and (\ref{LnB}), eliminating $B_{\mu\nu\alpha}$
in favor of $E_{\mu\nu}$, and keeping terms up to $O(\phi^2)$,
we obtain the desired second order equation for $E_{\mu\nu}$ as
\begin{eqnarray}
{\cal L}\, \hat{E}_{\mu\nu}
=\kappa_5^2\,S_{\mu\nu}\bigl[T^{(2)}_{ab}\bigr],
\label{2ndEeq}
\end{eqnarray}
where
\begin{eqnarray}
&& \hat{E}_{\mu\nu}=b^2(z) E_{\mu\nu},
\nonumber\\
&&{\cal L} \equiv b(z)\frac{\partial}{\partial z}\frac{1}{b(z)}
\frac{\partial}{\partial z} +\Box_4 -4H^2\,,
\nonumber   \\
&&S_{\mu\nu}\bigl[T^{(2)}_{ab}\bigr] \equiv b^4(z)
\Bigl[D^{\alpha}Q_{\alpha(\mu\nu)}\bigl[T^{(2)}_{ab} \bigr]
+{\pounds_{n}}\Bigl(P_{\mu\nu}
\bigl[T^{(2)}_{ab} \bigr] \Bigr)
+2\Bigl(\frac{b'(z)}{b^2(z)}\Bigr)P_{\mu\nu}
\bigl[T^{(2)}_{ab} \bigr]  +D_{(\mu}\Sigma_{\nu)}\bigl[T^{(2)}_{ab}\bigr]
\Bigr]\,,
 \nonumber \\
&&\Sigma_{\nu}\bigl[T^{(2)}_{ab}\bigr]
\equiv D^{\alpha}T^{\rm(b,2)}_{\nu\alpha}
+ 2\frac{b'(z)}{b^2(z)}
\Bigl(q^{\beta}_{\nu} T^{(2)}_{\beta\gamma} n^{\gamma}\Bigr)\,,
\label{defs}
\end{eqnarray}
where $T^{\rm(b,2)}_{\nu\alpha}$ is the second order part of
 $T^{\rm (b)}_{\nu\alpha}$ given by Eq.~(\ref{Tbmn}).
The operator ${\cal L}$ is the same as the one derived
in~\cite{Gen:2000nu}. What is new is that the source term
due to the presence of the bulk scalar field is taken into account.
The boundary condition on the brane is
\begin{eqnarray}
\frac{\partial}{\partial z}
\Bigl(\hat{E}_{\mu\nu} \Bigr)\Big|_{b}
=\kappa_5^2\,P_{\mu\nu}\bigl[T^{(2)}_{ab} \bigr]\big|_{b} \,,
 \label{Ebc}
\end{eqnarray}
which is obtained from Eq.~(\ref{LnE}) and
Israel's junction conditions~(\ref{israel}).
Thus, $E_{\mu\nu}$ has a jump at the brane in general,
whereas $\phi$ is smooth as given by Eq. (\ref{Neumann}).

In addition to the above, we have a set of constraint equations
that come from the 4-dimensional
contracted Bianchi identities on each $z=$constant hypersurface.
They are
\begin{eqnarray}
D^{\mu} E_{\mu\nu}=\kappa_5^2\,\Sigma_{\nu}\bigl[T^{(2)}_{ab}\bigr]\,.
\label{E4bianchi}
\end{eqnarray}
If we regard $E_{\mu\nu}$ as the `electric field',
these constraints imply that $\Sigma_{\nu}\bigl[T^{(2)}_{ab}\bigr]$
plays the role of the `charge density' in the bulk.

Although the bulk equation (\ref{2ndEeq}) is perfectly legitimate,
we find it is more convenient to deal with equations that
are reduced from Eq.~(\ref{2ndEeq}) by using the constraint
equation~(\ref{E4bianchi}), which we call
the reduced equations.

The components of Eq.~(\ref{2ndEeq}) are explicitly written down as
\begin{eqnarray}
&&b(z)\partial_{z} \Bigl(\frac{1}{b(z)}\partial_{z}\Bigr) \hatE_{t t}
+4H^2 \hatE_{tt}
- \partial_{t}^2 \hatE_{tt}-3 H \partial_{t} \hatE_{tt}
+\frac{1}{a^2}\nabla^2 \hatE_{tt}
-\frac{4H}{a^2}\partial^{k}\hatE_{k t}=\kappa_5^2\, S_{tt},
\label{Etteq}  \\
&&b(z)\partial_{z} \Bigl(\frac{1}{b(z)}\partial_{z}\Bigr) \hatE_{t i}
-2H^2 \hatE_{ti}
- \partial_{t}^2 \hatE_{ti}-H \partial_{t}\hatE_{ti}
+6 H^2 \hatE_{ti}
\nonumber\\
&&\hspace{6cm}+\frac{1}{a^2} \nabla^2 \hatE_{ti}
-2 H \partial_{i}\hatE_{tt}
-\frac{2 H}{a^2}\partial^{k}\hatE_{ki}=\kappa_5^2\,S_{t i},
\label{Etieq}\\
&&b(z)\partial_{z} \Bigl(\frac{1}{b(z)}\partial_{z}\Bigr) \hatE_{ij}
- \partial_{t}^2 \hatE_{ij}+H \partial_{t}\hatE_{ij}
\nonumber\\
&&\hspace{4cm}+\frac{1}{a^2} \nabla^2 \hatE_{ij}
-2 H(\partial_{i}\hatE_{tj}+\partial_{j}\hatE_{ti})
+2 a^2 H^2\delta_{i j}\hatE_{tt}=\kappa_5^2\,S_{ij}\,.
\label{Eijeq}
\end{eqnarray}
In the above equations, the components of $E_{\mu\nu}$ are
coupled each other, which makes it very difficult to solve
them.

However, this situation can be improved by using the
constraint equation~(\ref{E4bianchi}).
The components of Eq.~(\ref{E4bianchi}) are written down as
\begin{eqnarray}
&&-\partial_{t}\hatE_{tt}+\frac{1}{a^2}\partial^{k}
\hatE_{k t}-4 H\hatE_{tt}
=\kappa_5^2\,b^4(z) \Sigma_{t}\,,
\label{Ettconstr} \\
&&-\partial_{t}\hatE_{ti}+\frac{1}{a^2}\partial^{k}\hatE_{k i}
-3 H\hatE_{ti}
=\kappa_5^2\,b^4(z)\Sigma_{i}\,.
\label{Eticonstr}
\end{eqnarray}
Using these, we may eliminate $\partial^k\hat E_{kt}$ and
$\partial^k\hat E_{ki}$ from Eqs.~(\ref{Etteq}) and (\ref{Etieq}),
respectively, to obtain
\begin{eqnarray}
&&b(z)\partial_{z} \Bigl(\frac{1}{b(z)}\partial_{z}\Bigr) \hatE_{t t}
-12H^2 \hatE_{tt}
- \partial_{t}^2 \hatE_{tt}-7 H \partial_{t} \hatE_{tt}
+\frac{1}{a^2}\nabla^2 \hatE_{tt}
=\kappa_5^2\,\tilde{S}_{tt}\,;
\nonumber\\
&&\hspace{7cm}\tilde{S}_{tt}=S_{tt}+4 H b^4(z) \Sigma_{t}\,,
\label{Ettreq} \\
&&b(z)\partial_{z} \Bigl(\frac{1}{b(z)}\partial_{z}\Bigr) \hatE_{t i}
-2H^2 \hatE_{ti}
- \partial_{t}^2 \hatE_{ti}-3 H \partial_{t}\hatE_{ti}
+\frac{1}{a^2} \nabla^2 \hatE_{ti}
=\kappa_5^2\,\tilde{S}_{ti}\,;
\nonumber\\
&&\hspace{7cm}
\tilde{S}_{ti}
=S_{t i}+2 Hb^4(z) \Sigma_{i}+\frac{2H}{\kappa_5^2}\,
 \partial_{i}\hatE_{tt}\,,
\label{Etireq}
\end{eqnarray}
and Eq.~(\ref{Eijeq}) is rewritten as
\begin{eqnarray}
&&b(z)\partial_{z} \Bigl(\frac{1}{b(z)}\partial_{z}\Bigr) \hatE_{ij}
-\partial_{t}^2 \hatE_{ij}+H \partial_{t}\hatE_{ij}
+\frac{1}{a^2} \nabla^2 \hatE_{ij}
=\kappa_5^2\,\tilde{S}_{ij}\,;
\nonumber\\
&&\hspace{5cm}
\tilde{S}_{ij}=S_{ij}
+\frac{1}{\kappa_5^2}
\left( 2H(\partial_{i}\hatE_{tj}+\partial_{j}\hatE_{t i})
-2 a^2 H^2\delta_{ij}\hatE_{tt}\right)\,.
\label{Eijreq}
\end{eqnarray}
Thus, one can determine the components of $E_{\mu\nu}$ step by step;
solving first Eq.~(\ref{Ettreq}) for $E_{tt}$,
next Eq.~(\ref{Etireq}) for $E_{ti}$,
and finally Eq.~(\ref{Eijreq}) for $E_{ij}$.

\subsection{The spatially homogeneous and isotropic background}

Here, we solve the evolution equation~(\ref{Ettreq}) in the
spatially homogeneous and isotropic case. In this case, we have already
seen that $E_{\mu\nu}$ on the brane can be obtained without
solving the bulk, as given by Eq. (\ref{Ettlate}).
The purpose of this subsection is to determine
$E_{\mu\nu}$ in the bulk and to take the limit of it to the brane to
examine if $E_{\mu\nu}$ thus obtained gives the correct answer
for $E_{\mu\nu}$ on the brane, as a check of our derivation given above.

Equation~(\ref{Ettreq}) in the spatially homogeneous case is
\begin{eqnarray}
\Bigl[b(z)\partial_{z} \Bigl(\frac{1}{b(z)}\partial_{z}\Bigr)
-\partial_{t}^2-7 H \partial_{t}
-12 H^2 \Bigr]\hatE_{tt}
=\kappa_5^2\, \tilde{S}_{tt}\,.
\label{Etthomoeq}
\end{eqnarray}
To solve this, we introduce the retarded Green function
that satisfies
\begin{eqnarray}
\Bigl[b(z)\partial_{z}\Bigl(\frac{1}{b(z)}\partial_{z}\Bigr)
 -\partial_{t}^2 - 7 H\partial_{t} -12H^2 \Bigr]
 G(t,z;t',z')
=-\delta(t-t')\delta(z-z')b(z).
\label{GretEtt}
\end{eqnarray}
The Green function satisfies the reciprocity condition,
\begin{eqnarray}
G(t,z;t',z')=G(-t',z';-t,z),
\label{Grecip}
\end{eqnarray}
and the Neumann-type boundary condition on the brane,
\begin{eqnarray}
\partial_{z} G(t,0+;t',z')=0\,.
\label{Gretbc}
\end{eqnarray}
An explicit construction of the retarded Green function is given in
Appendix~\ref{Greenfcn}.

Using the Green function, the formal solution of $E_{tt}$ is
expressed as
\begin{eqnarray}
\hat{E}_{tt}(t,z)
&=&-\int_0^\infty dt' \frac{1}{b(z')}
G(t,z;t',z')\partial_{z'}\hat{E}_{tt}(t',z')\Bigg|_{z'=0}
\nonumber\\
&&-\int_0^\infty\frac{dz'}{b(z')}
\Bigl[\partial_{t'}G(t,z;t',z')
-G(t,z;t',z') \partial_{t'}
+7H G(t,z;t',z')\Bigr]\hat{E}_{tt}(t',z')\Bigg|_{t'=0}
\nonumber\\
&&-\kappa_5^2\int_0^\infty dt'\int_0^\infty dz' \frac{1}{b(z')} G(t,z;t',z')
\tilde{S}_{tt}(t',z')\,,
\label{Ettfsol}
\end{eqnarray}
where we have assumed a regular initial data at $t=0$ so that
there is no contribution from $z=\infty$.
In the above,
the first term is the contribution from the brane-boundary of the
extra-dimension, the second term is that from the initial
surface, and the third term is that from the source in the bulk
generated from the scalar field.
In Appendix~\ref{Greenfcn},
we show that the Green function $G(t,z;t',z')$
at late times behaves as $\propto a(t)^{-4}$. Therefore, the second term
behaves as radiation, i.e., it describes the dark radiation term.
Thus for a de Sitter brane, or for any observer on the $z=$constant
hypersurface, we can neglect it after a sufficient lapse of time.

The third term can be evaluated as follows.
Using the field equation~(\ref{phieq}),
we find the relations among several terms in the source term as
\begin{eqnarray}
&&D^{\alpha}Q_{\alpha tt}+\frac{\dot{a}}{a}\Sigma_{t}=0 ,
\label{DQ}
\\
&&\partial_t \Bigl(a^4b^3 P_{tt}\Bigr)
= -\partial_{z}\Bigl(a^4b^4\Sigma_{t} \Bigr).
\label{dtP}
\end{eqnarray}
In particular, the latter equality implies the existence of
an integrability condition.
Namely, if we define
\begin{eqnarray}
f(t,z)\equiv a^4(t) b^3(z) P_{tt}\,,
\quad
g(t,z)\equiv -a^4(t) b^4(z) \Sigma_{t}\,,
\label{fgdef}
\end{eqnarray}
Eq.~(\ref{dtP}) implies the existence of a function $M(t,z)$
such that
\begin{eqnarray}
f(t,z)=\partial_{z}M(t,z)\,,
\quad
g(t,z)= \partial_{t}M(t,z).
\label{fgform}
\end{eqnarray}
Introducing a rescaled function of $M(t,z)$ as
\begin{eqnarray}
\tilde{M}(t,z)\equiv a^{-4}(t) M(t,z),
\label{tildeM}
\end{eqnarray}
the source function $\tilde{S}_{tt}$ is expressed as
\begin{eqnarray}
\tilde{S}_{tt}(t,z)= \Bigl[b(z)\partial_{z}
\Bigl(\frac{1}{b(z)}\partial_{z}
\Bigr)-\partial_{t}^2 -7 H\partial_{t} -12H^2\Bigr]\tilde{M}(t,z).
\label{tildeSexp}
\end{eqnarray}
We see that the operator acting on $\tilde{M}(t,z)$ is the same as
that acting on the Green function in Eq. (\ref{GretEtt}).
Therefore, by repeating integration by part twice,
the third term of Eq.~(\ref{Ettfsol}) can be
rewritten in the form,
\begin{eqnarray}
\int_0^\infty dt' \frac{1}{b(z')}
G(t,z;t',z')\partial_{z'}\tilde{M}(t',z')\Bigg|_{z'=0}
+\tilde{M}(t,z)\,.
\label{bulkterm}
\end{eqnarray}
Here again, assuming the regularity of the scalar field, we have
neglected the contribution from $z=\infty$.
The boundary condition of $\tilde{M}(t,z)$ on the brane-boundary
is obtained from its definition as
\begin{eqnarray}
\partial_{z} \tilde{M}(t,0+)=P_{tt}(t,0+).
\label{tildeMbc}
\end{eqnarray}
This is the same as the boundary condition for $\hat{E}_{tt}$,
Eq.~(\ref{Ebc}).
Thus the boundary contribution in Eq.~(\ref{bulkterm}) just cancels
the first term in Eq.~(\ref{Ettfsol}),
and we have
\begin{eqnarray}
\hat{E}_{tt}(t,z)
=\kappa_5^2\,\tilde{M}(t,z)\,,
\label{Etthomosol}
\end{eqnarray}
that is, the solution for $\hat{E}_{tt}$ is given by $\tilde{M}$,
apart from the initial surface contribution that can be neglected
at late times.

{}From Eqs.~(\ref{fgform}) and (\ref{tildeM}),
it immediately follows that two different representations of $\tilde{M}$,
 hence of $\hat{E}_{tt}$, are possible;
the $t$-integral representation and the $z$-integral representation.
\begin{enumerate}

\item{$t$-integral representation}

The $t$-integral representation of the solution is
\begin{eqnarray}
E_{tt}(t,z)=-\kappa_5^2\,
\frac{b^2(z)}{a(t)^4} \int^{t}dt' a^4(t')
 \Bigl(D^{\mu}T^{\rm(b,2)}_{\mu t}(t',z)
+2\frac{b'(z)}{b(z)}\,T^{(2)}_{nt}(t',z) \Bigr)\,.
\label{tintrep}
\end{eqnarray}
Taking the limit of it to the brane,
and eliminating the mass term using the field equation,
we obtain
\begin{eqnarray}
E_{tt}(t,0+)=\frac{1}{a^4(t)}
\left[\frac{\kappa_5^2}{2}\int^{t}_{0} dt' a^4(t')
\Bigl(\partial_{z}^2\phi+\frac{\dot{a}}{a}\dot{\phi} \Bigr) \dot{\phi}
+a^4(0)E_{tt}(0,0+)\right]\,,
\label{Ettbrane}
\end{eqnarray}
where we have recovered the initial surface contribution.
This is the same as Eq.(\ref{Ettlate})
obtained from the contracted Bianchi identities on the brane.

\item{$z$-integral representation}

The $z$-integral representation of the solution is
\begin{eqnarray}
E_{tt}(t,z)=\frac{1}{b(z)^{2}}\left[
\kappa_5^2\int_0^z dz'\,b^3(z') P_{tt}(t,z')
+E_{tt}(t,0+)\right].
\label{zintrep}
\end{eqnarray}
This is a new form that has not been obtained before.
A nice feature of this representation
is that it automatically satisfies the
boundary condition on the brane.
One immediate consequence is that the regularity at $z=\infty$
implies the equality,
\begin{eqnarray}
E_{tt}(t,0+)=-\kappa_5^2\int_0^\infty dz\,b^3(z) P_{tt}(t,z)\,.
\label{Ettzint}
\end{eqnarray}
Thus, $E_{tt}$ on the brane may be expressed in terms of
an integral of over the extra-dimension on the $t=$constant
hypersurface with the integrand given by a certain combination of
second derivatives of the energy momentum tensor.
A point to be emphasized is that Eq.~(\ref{Ettzint})
has no explicit dependence on the initial condition;
$E_{tt}$ on the brane is solely determined
by the field configuration on the $t=$ constant hypersurface.
This is similar to the local mass conservation law
that holds in a spherically symmetric spacetime.
Apparently, Eqs.~(\ref{Ettbrane}) and (\ref{Ettzint}) must
coincide with each other up to the term proportional to $a^{-4}(t)$.
This is guaranteed by Eq.~(\ref{dtP}). 
The representation~(\ref{zintrep}) may be useful
for analyzing the behavior of $E_{tt}$ in the bulk, though
we have not explored it yet.

\end{enumerate}

Given the solution for $E_{tt}$, it is trivial to
obtain the other components $E_{ti}$ and $E_{ij}$.
For $E_{ti}$ and the traceless part of $E_{ij}$, the
source terms for a spatially homogeneous $\phi$ are easily found
to be zero. Hence they are zero except for the contributions
from the initial data that decay rapidly in time anyway.
The trace part of $E_{ij}$ is simply given by $E_{tt}$
because of the traceless nature of $E_{\mu\nu}$.

\section{Cosmological Perturbations}

 To investigate cosmological perturbations on the brane,
we take the following approach.
First we derive the perturbation equations for the scalar field $\delta\phi$
and the projected Weyl tensor $\delta E_{\mu\nu}$ in the bulk.
Then we solve them with the assumption that the scale of interest is
much greater than the Hubble horizon size, and
take their projections on the brane to evaluate their effects.

Note that there are two small parameters in our approach. As before,
we assume the amplitude of $\phi$ to be small. In addition to it,
we assume $\delta\phi$ is much smaller than $\phi$, say
$\delta\phi=O(\epsilon\phi)$ with $\epsilon\ll1$. In what follows,
we consider the perturbations accurate to $O(\phi^2)$ in the scalar
field amplitude and linear in $\epsilon$.
An important consequence is that
the perturbation of the energy momentum tensor becomes effectively
gauge-invariant as will be shown below. Our calculations are
considerably simplified by this fact.

We focus on the so-called scalar-type cosmological perturbations,
because the tensor-type perturbations (which are spatially
transverse-traceless on the brane) are identical to those
in the vacuum AdS$_5$ bulk model discussed in the
literature~\cite{Garriga:1999bq,Langlois:2000ns}
to the accuracy of $O(\phi^2)$.

\subsection{The perturbation equations in the bulk}

First, we write down the perturbation equations in the bulk.
As usual, we expand all the perturbation variables in terms of
the spatial scalar harmonics $Y(x^i)$ that satisfy
\begin{eqnarray}
\left(\delta^{ij}\partial_i\partial_j+k^2\right)Y=0\,,
\end{eqnarray}
and the associated vector and tensor harmonics
\begin{eqnarray}
Y_i=-{1\over k}\partial_i Y\,,
\quad
Y_{ij}={1\over k^2}\partial_i\partial_jY+{1\over3}\delta_{ij}Y\,,
\end{eqnarray}
where we have assumed spatially flat slicing of each $z=$constant
hypersurface for simplicity.

The perturbation of the scalar field equation is simply given by
\begin{eqnarray}
\left[{1\over b^3}\partial_zb^3\partial_z
-{1\over a^3}\partial_ta^3\partial_t
-{k^2\over a^2}-m^2b^2\right]\chi(t,z)=0\,,
\label{deltaphieq}
\end{eqnarray}
with the boundary condition $\partial_z\chi|_{b}=0$,
where we have set $\delta\phi=\chi(t,z)Y(x^i)$.
Note that we do not need to take account of the metric
perturbation in the bulk because it will be of $O(\phi^2)$.

To write down the perturbation equations for $\delta E_{\mu\nu}$,
we expand $\delta E_{\mu\nu}$ in terms of
the scalar harmonics as
\begin{eqnarray}
&&\delta\hat{E}_{tt}= E Y\,,
\nonumber\\
&&\delta\hat{E}_{ti}=aE_{1}Y_{i}\,,
\nonumber\\
&&\delta\hat{E}_{ij}=a^2\bigl(\frac{1}{3}EY\delta_{ij}
+E_{2}Y_{ij} \bigr)\,,
\label{deltaE}
\end{eqnarray}
and their source terms
$S_{\mu\nu}[\delta T^{(2)}_{ab}]$ and
$\Sigma_{\nu}[\delta T^{(2)}_{ab}]$
as
\begin{eqnarray}
&&S_{tt}[\delta T^{(2)}_{ab}]= S Y\,,
\nonumber\\
&&S_{ti}[\delta T^{(2)}_{ab}]=aS_{1}Y_{i}\,,
\nonumber\\
&&S_{ij}[\delta T^{(2)}_{ab}]=a^2
\bigl(\frac{1}{3}SY\delta_{ij}+S_{2}Y_{ij} \bigr)\,,
\label{deltaS}
\\
&&\Sigma_{t}[\delta T^{(2)}_{ab}]=\Sigma Y\,,
\nonumber\\
&&\Sigma_{i}[\delta T^{(2)}_{ab}]=a\Sigma_1 Y_{i}\,.
\label{deltaSig}
\end{eqnarray}

The evolution equations of $\delta E_{\mu\nu}$ are written down as
\begin{eqnarray}
&&\Bigl[b\partial_{z}b^{-1}\partial_{z}
-a^{-7}\partial_ta^{7}\partial_t
-12H^2-{k^2\over a^2} \Bigr]E=
\kappa_5^2\,S+4Hb^4\Sigma\,,
\nonumber\\
&&\Bigl[b\partial_{z}b^{-1}\partial_{z}
-a^{-5}\partial_ta^5\partial_t
-6H^2-{k^2\over a^2}\Bigr]E_1
=\kappa_5^2\left(S_1+2Hb^4\Sigma_{1}\right)-2{k\over a}H E\,,
\nonumber \\
&&\Bigl[b\partial_{z}b^{-1}\partial_{z}
-a^{-3}\partial_ta^3\partial_t
-2H^2-{k^2\over a^2}\Bigr]E_2
=\kappa_5^2\, S_2-4{k\over a}HE_1\,,
\label{Eeq}
\end{eqnarray}
and the constraint equations are written as
\begin{eqnarray}
&&-\Bigl(\partial_{t}+4H \Bigr)E+{k\over a}E_{1}
=\kappa_5^2\,b^4\Sigma\,,
\nonumber\\
&&-\Bigl(\partial_{t}+4H \Bigr)E_1
-\frac{k}{3a}\bigl(E-2E_2\bigr)
=\kappa_5^2\,b^4\Sigma_{1}\,.
 \label{Econstr}
\end{eqnarray}

Using the expression of $S_{\mu\nu}[\delta T^{(2)}_{ab}]$ given
in Eq.~(\ref{defs}), we may further deduce the source term as
follows. We decompose $P_{\mu\nu}[\delta T^{(2)}_{ab}]$
as
\begin{eqnarray}
&&P_{tt}[\delta T^{(2)}_{ab}]=P Y,\nonumber\\
&&P_{ti}[\delta T^{(2)}_{ab}]=aP_{1}Y_{i},\nonumber\\
&&P_{ij}[\delta T^{(2)}_{ab}]
=a^2\bigl(\frac{1}{3}PY\delta_{ij}+P_{2}Y_{ij} \bigr),
\label{deltaP}
\end{eqnarray}
and $D^{\alpha}Q_{\alpha(\mu\nu)}[\delta T^{(2)}_{ab}]$ as
\begin{eqnarray}
&&D^{\alpha}Q_{\alpha tt}[\delta T^{(2)}_{ab}]=Q Y,
\nonumber\\
&&D^{\alpha}Q_{\alpha(ti)}[\delta T^{(2)}_{ab}]=a Q_{1}Y_{i},
\nonumber\\
&&D^{\alpha}Q_{\alpha (ij)}[\delta T^{(2)}_{ab}]
=a^2\bigl(Q_{0}Y\delta_{ij}+Q_{2}Y_{ij} \bigr),
\label{deltaDQ}
\end{eqnarray}
where, from the traceless condition of the whole source term, $Q_{0}$ is
expressed as
\begin{eqnarray}
Q_0=\frac{1}{3}\Bigl[Q+a^{-3}\partial_{t}(a^3\Sigma)
-\frac{k}{a}\Sigma_{1} \Bigr].
\label{Qzero}
\end{eqnarray}
With these expressions, $S$, $S_{1}$ and $S_{2}$ are
expressed as
\begin{eqnarray}
&&S=b^4\Bigl[b^{-3}\partial_z(b^2P)+Q
+\partial_{t}\Sigma\Bigr],
\nonumber\\
&&S_1=
b^4\Bigl[b^{-3}\partial_z(b^2P_1)+Q_1
-\frac{1}{2}\Bigl\{\frac{k}{a}\Sigma
-a\partial_t(a^{-1}\Sigma_{1})\Bigr\}\Bigr]
\nonumber\\
&&S_{2}=
b^4\Bigl[b^{-3}\partial_z\Bigl(b^2P_2\Bigr)+Q_2
-\frac{k}{a}\Sigma_{1}\Bigr].
\label{Sform}
\end{eqnarray}
The explicit forms of these source terms in terms of the
scalar field perturbation are given in Appendix~\ref{Source}.

The boundary conditions of $\delta E_{\mu\nu}$ on the
brane are
\begin{eqnarray}
\partial_{z} E=\kappa_5^2\,P,
\quad
\partial_{z} E_1=\kappa_5^2\,P_1,
\quad
\partial_{z} E_2=\kappa_5^2\,P_2,
\label{pertbc}
\end{eqnarray}
by using Eq. (\ref{Ebc}).

The bulk effects on the brane can be analyzed by
solving first the scalar field perturbation equation (\ref{deltaphieq}),
inserting it to the source term of Eq. (\ref{Eeq}),
and solving it under the boundary conditions (\ref{pertbc}).
Although to solve them analytically is quite difficult or almost impossible,
it is possible to solve them in the long wavelength limit $k/aH \ll1$.

\subsection{$\bm{\delta\phi}$ in the long wavelength limit}

Let us analyze the behavior of the scalar field perturbation $\chi$
in the long wavelength limit. We expand it as
\begin{eqnarray}
\chi(t,z)=\chi_{b}(t,z)
+\int_{3H/2}^\infty dM \,\chi_{M} (t,z),
\label{chidecomp}
\end{eqnarray}
where the first term denotes the contribution from the bound state mode
with the effective 4-dimensional mass-squared $M^2_{\rm eff}=m^2/2$
(under the assumption $m^2\ll H^2$),
and the second term is a superposition of the continuous
Kaluza-Klein (KK) modes. Since the KK modes have mass-squared
$M^2$ larger than $9H^2/4$, once their wavelengths become larger
than the Hubble horizon size, they will damp out exponentially
as $a^{-3/2}$. Hence we expect that their contributions to be
unimportant.
At least in the limit $H\ell\ll1$, there are many pieces of
evidence that support this expectation.
As in the conventional inflationary scenario,
the scalar field perturbations will come from the quantum
fluctuations, and this expectation should be carefully
examined in the context of quantum theory.
Here, however, we simply assume so and focus on the bound state mode.
Then we have the relation,
\begin{eqnarray}
{1\over b^{3}}\partial_z(b^3\partial_z\chi)
=\left(m^2b^2-{m^2\over 2}\right)\chi=0\,.
\end{eqnarray}
In particular, on the brane, this gives
\begin{eqnarray}
\partial_z^2\chi\big|_b={m^2\over2}\chi
=-{1\over a^{3}}\partial_t(a^3\partial_t\chi)-{k^2\over a^2}\chi\,.
\label{d2chi}
\end{eqnarray}

\subsection{$\bm{\delta E_{\mu\nu}}$ in the long wavelength limit}

By inspecting the explicit forms of the source terms $S$, $S_1$,
$S_2$, $\Sigma$ and $\Sigma_1$, we find their behaviors as
\begin{eqnarray}
S,\,\Sigma=O(1),\quad
S_1\,,\Sigma_1=O(k),\quad
S_2=O(k^2)\quad\mbox{for}\quad k\to0.
\end{eqnarray}
Then from Eqs.~(\ref{Eeq}) we find
\begin{eqnarray}
E=O(1),\quad
E_1=O(k),\quad
E_2=O(k^2).
\end{eqnarray}
Turning to the constraint equations (\ref{Econstr}), we then see
that the long wavelength limit of $E$ and $E_1$ can be obtained
by simply integrating Eqs.~(\ref{Econstr}), i.e., without
solving Eqs.~(\ref{Eeq}), while we need to solve the
2nd order bulk equation to obtain $E_2$.
Therefore, let us first solve for $E$ and $E_1$.
Discussion on $E_2$ is deferred to the next subsection.

At leading order in the long wavelength limit,
Eqs.~(\ref{Econstr}) reduce to
\begin{eqnarray}
&&-{1\over a^4}\partial_{t}(a^4E)
=\kappa_5^2\,b^4\Sigma+O(k^2),
\label{Elimit}\\
&&-{1\over a^4}\partial_{t}(a^4E_1)
=\kappa_5^2\,b^4\Sigma_{1}+\frac{k}{3a}E+O(k^3),
\label{Eonelimit}
\end{eqnarray}
Thus $E$ is given by
\begin{eqnarray}
E(t,z)=-\kappa_5^2\,
{b^4(z)\over a^4(t)}\int^tdt'a^4(t')\Sigma(t',z)+O(k^2)\,,
\label{Elong}
\end{eqnarray}
where
\begin{eqnarray}
\Sigma&=&{1\over 2b^2}\Biggl[
{b'\over b}(\phi'\dot\chi+\dot\phi\chi')
+(\phi'\dot\chi{}'+\dot\phi{}'\chi')
-(\phi''\dot\chi+\dot\phi\chi'')-2H\dot\phi\dot\chi\Biggr]\,.
\label{Sigbulk}
\end{eqnarray}
With thus given $E$, $E_1$ is given by
\begin{eqnarray}
E_1(t,z)=-{b^4(z)\over a^4(t)}
\int^tdt'a^4(t')\left(\kappa_5^2\,
\Sigma_1+{k\over3\kappa_{5}^2ab^4}E\right)(t',z)+O(k^3)\,,
\label{E1long}
\end{eqnarray}
where
\begin{eqnarray}
\Sigma_1=-{1\over 2b^2}{k\over a}
\Biggl[-m^2b^2\phi\chi+4{b'\over b}\phi'\chi
-4H\dot\phi\chi-{4\over3}\ddot\phi\chi
+\phi'\chi'+{1\over3}\dot\phi\dot\chi\Biggr]\,.
\label{Sig1bulk}
\end{eqnarray}

On the brane, the time integrals in Eqs.~(\ref{Elong}) and (\ref{E1long})
can be explicitly performed. Using Eq.~(\ref{d2chi}) and the boundary
condition $\phi'=\chi'=0$,
$\Sigma$ on the brane can be rewritten as
\begin{eqnarray}
\Sigma(t,0+)
&=&{1\over2}(\ddot\phi\dot\chi+\dot\phi\ddot\chi)
+2H\dot\phi\dot\chi+{k^2\over2a^2}\dot\phi\chi
\nonumber\\
&=&
{1\over2a^4}\left(a^4\dot\phi\dot\chi\right)\!\dot{\,}
+{k^2\over2a^2}\dot\phi\chi\,.
\end{eqnarray}
Hence we obtain
\begin{eqnarray}
E(t,0+)=-{\kappa_5^2\over2}\,
\dot\phi\dot\chi+{C\over a^4}+O(k^2)\,,
\label{Ebrane}
\end{eqnarray}
where $C$ is an integration constant. Since the term $C/a^4$
decays rapidly, we may neglect it at late times.
Then, with a similar manipulation, we find
\begin{eqnarray}
\left(\kappa_{5}^2\,\Sigma_1+{k\over3a}E\right)(t,0+)
=-{\kappa_5^2\over3}{k\over a}\left(
\ddot\phi\chi+\dot\phi\dot\chi+3H\dot\phi\chi\right)
=-{\kappa_5^2\over3}{k\over a^4}\left(a^3\dot\phi\chi\right)\!\dot{\,}\,,
\end{eqnarray}
which gives
\begin{eqnarray}
E_1(t,0+)={\kappa_5^2\over3}{k\over a}\dot\phi\chi\,,
\label{E1brane}
\end{eqnarray}
where we have neglected the term $\propto a^{-4}$ as before.
Thus we have obtained $E$ and $E_1$ on the brane on superhorizon
scales.

\subsection{Bulk anisotropic effect}

Now we turn to solving $E_2$ that describes the spatially anisotropic
part of $\delta E_{\mu\nu}$.
In the long wavelength limit, the equation for $E_{2}$ in Eqs.~(\ref{Eeq})
reduces to
\begin{eqnarray}
\Bigl[b\partial_{z}\Bigl(\frac{1}{b}\partial_{z}\Bigr)
-{1\over a^3}\partial_{t}\Bigl(a^3\partial_{t}\Bigr)
-2H^2 \Bigr]E_{2}=\kappa_5^2\,S_{2}-4\frac{k}{a}HE_{1}+O(k^4).
\label{E2longeq}
\end{eqnarray}
The $E_1$ term on the right-hand side of the above equation is
given by the time integral in Eq.~(\ref{E1long}). In general,
it is impossible to perform the integral explicitly. However,
if we assume that $\phi$ is dominated by the bound-state mode,
its time variation is small; i.e., we may adopt the slow-roll
approximation. We have already assumed that the perturbation
$\chi$ is dominated by the bound-state mode as well.
Hence, under the assumption $m^2\ll H^2$, $E_1$
to the lowest order in $m^2$ is found as
\begin{eqnarray}
E_1\approx -{\kappa_5^2\over6}\,b^2{k\over aH}
\left(\phi''\chi+H\dot\phi\chi-{b'\over b}\phi'\chi\right).
\label{E1slow}
\end{eqnarray}
Note, however, that this approximation will not be
valid at large values of $H|z|$, since the
regularity of the spacetime at $H|z|=\infty$
implies the breakdown of the bound-state mode dominance.
Nevertheless, it seems physically reasonable to assume
that the precise behavior of the source term of $E_2$
at large values of $H|z|$ will not significantly
affect the value of $E_2$ on the brane.
Under the same approximation, $S_2$ is given by
\begin{eqnarray}
S_2\approx-{b^2\over3}\,{k^2\over a^2}
\left(4\phi''\chi+2H\dot\phi\chi\right)\,.
\label{S2slow}
\end{eqnarray}
{}From Eqs.~(\ref{E1slow}) and (\ref{S2slow}),
the source term is evaluated as
\begin{eqnarray}
\kappa_5^2\,S_2-4H{k\over a}E_1
\approx
-\kappa_5^2\,{2b^2\over3}\,{k^2\over a^2}
\left(\phi''\chi+{b'\over b}\phi'\chi\right).
\label{E2ss}
\end{eqnarray}

Given the source term, the formal solution of $E_{2}$ is
given by
\begin{eqnarray}
E_{2}(t,z)=-\int_{0}^{t}dt' \int_{0}^{\infty}dz'
\frac{1}{b(z')} G_{2}(t,z;t',z')
\Bigl(\kappa_5^2\,S_{2}-4H\frac{k}{a}E_{1}\Bigr)(t',z'),
\label{E2fsol}
\end{eqnarray}
where $G_2$ is the retarded Green function for $E_2$
that satisfies
\begin{eqnarray}
\Bigl[b\partial_{z}b^{-1}\partial_{z}
-a^{-3}\partial_ta^3\partial_t-2H^2\Bigr]G_2(t,z;t',z')
= -b(z)\delta(t-t')\delta(z-z')\,,
\end{eqnarray}
in the long wavelength limit.
In the above, we have neglected the contribution from the initial
surface $t=0$ since $\phi$ is assumed to be very near the top
of the potential at $t=0$. The contribution from the brane-boundary is
absent because $P_{2}=0$ on the brane.
For $Hl\ll1$, and at late times, the retarded Green function
behaves as
\begin{eqnarray}
G_{2}(t,z;t',z')=(Hl)\theta(t-t') \,a^{-2}(t)\,a^{2}(t')\,,
\label{G2late}
\end{eqnarray}
as shown in Appendix~\ref{Greenfcn}.

Then, at late times, the $z$-integral in Eq.~(\ref{E2fsol})
can be easily evaluated. From Eq.~(\ref{E2ss}) we find
\begin{eqnarray}
\int_0^\infty dz{1\over b(z)}
\Bigl(\kappa_5^2\,S_{2}-4H\frac{k}{a}E_{1}\Bigr)(t,z)
&\propto&
\int_0^\infty dz\left(b\phi''+b'\phi'\right)\chi
\nonumber\\
&=&\int_0^\infty dz\left[\left(b\phi'\chi\right)'-b\phi'\chi'\right]
=-\int_0^\infty dz\,b\phi'\chi'\,.
\end{eqnarray}
Now, under the assumptions that $\phi$ and $\chi$ are both
dominated by the bound-state mode and $m^2\ell^2\ll H^2\ell^2\ll1$,
the derivatives $\phi'$ and $\chi'$ are both of
$O(m^2)$.
Hence $E_2$ turns out to be of $O(m^4)$.
This is a very interesting result.
As long as we focus on the effects that
persist after a sufficiently long lapse of time,
signatures of the braneworld in this
bulk inflaton model do not appear at $O(m^2)$ but
only at $O(m^4)$. Since we have neglected $O(m^4)$ terms
when deriving the source term (\ref{E2ss}),
we are not able to evaluate the $O(m^4)$ corrections
quantitatively here, but leave it for future work.
It may be worth mentioning that analyzes of
the decay of the scalar field out to the bulk were carried
out previously and this effect was shown to appear at
the same order, i.e.,
at $O(m^4)$~\cite{Dubovsky:2000am,Himemoto:2001hu,Langlois:2003dd}.

\subsection{Cosmological perturbations on the brane}

We are now ready to discuss cosmological perturbations on the brane.
General behavior of superhorizon scale cosmological perturbations on
the brane has been analyzed with the geometrical approach
in~\cite{Langlois:2000iu}.
Here we follow their analysis.
Since the scalar field perturbation induces only scalar-type
perturbations on the brane, the perturbed metric on the brane
$\gamma_{\mu\nu}=q_{\mu\nu}/b^2(0+)$ may be expressed as
\begin{eqnarray}
&&\gamma_{tt}=-\bigl(1+2A\,Y\bigr),
\nonumber\\
&&\gamma_{ti}=-a(t)B\,Y_{i},
\nonumber\\
&&\gamma_{ij}=
a^2(t)\bigl(\delta_{ij}+2H_{L}\,Y\delta_{ij}+2H_{T}\,Y_{ij}\bigr)\,.
\label{pertmetric}
\end{eqnarray}
The effective energy-momentum tensor on the brane is
\begin{eqnarray}
T^{({\rm eff})}_{\mu\nu}=
\ell_0\left(T^{{\rm (b)}}_{\mu\nu}
-\frac{1}{\kappa_{5}^2}E_{\mu\nu}\right),
\label{effTmn}
\end{eqnarray}
where $T^{\rm (b)}_{\mu\nu}$ is given by Eq.~(\ref{Tbmn}).
The perturbed effective energy-momentum tensor
$\delta T^{({\rm eff})}_{\mu\nu}$ is denoted by
\begin{eqnarray}
&&\delta T^{({\rm eff})t}_{\qquad t}=-\rho\, \delta \,Y,
\nonumber\\
&&\delta T^{({\rm eff})t}_{\qquad i}=q\, Y_{i},
\nonumber\\
&&\delta T^{({\rm eff})i}_{\qquad j}
=p\bigl(\pi_{L}\, Y \,\delta^{i}_{j}+\Pi \,Y^{i}_{j}\bigr),
\label{pertTmn}
\end{eqnarray}
where $\delta$, $q$, $\pi_{L}$ and $\Pi$ represent the density
contrast, momentum density fluctuation, isotropic
pressure perturbation and anisotropic stress perturbation,
 respectively.
By taking the perturbation of $T^{\rm(b)}_{\mu\nu}$ and $E_{\mu\nu}$
explicitly, these matter perturbation variables are
expressed in terms of $\delta\phi$ and $\delta E_{\mu\nu}$ as
\begin{eqnarray}
&&\rho\,\delta=\ell_0
\bigl({1\over2}\dot\phi \,\dot\chi
+{1\over2}m^2 \phi\,\chi-\frac{1}{\kappa_{5}^2}E\bigr),
\nonumber\\
&&q=\ell_0\bigl({2\over3}{k\over a}\dot\phi\,\chi
+\frac{1}{\kappa_{5}^2}E_{1} \bigr) ,
\nonumber\\
&&p\,\pi_{L}=\ell_0
\bigl({5\over6}\dot\phi \,\dot\chi
-{1\over2}m^2 \phi\,\chi-\frac{1}{3\kappa_{5}^2}E\bigr),
\nonumber\\
&&p\,\Pi=-\frac{\ell_0}{\kappa_{5}^2} E_{2},
\nonumber\\
\end{eqnarray}
where it is understood that all the quantities are
those evaluated on the brane.

We have just seen in subsection D that $E_2$ appears only at $O(m^4)$.
Since $m^2/H^2\ll 1$ is assumed,
this implies that anisotropic stress is of $O(m^4\ell^4)$
in the low energy approximation $H^2\ell^2\ll1$.
Hence, to the leading order in the low energy expansion,
the anisotropic stress perturbation vanishes; $\Pi=0$.
The other matter perturbation variables are expressed
in terms of $\chi$ by substituting to the above the expressions of
$E$ and $E_1$, Eqs.~(\ref{Ebrane}) and (\ref{E1brane}), respectively.
We thus find
\begin{eqnarray}
&&\rho\,\delta=\ell_0
\bigl(\dot\phi \,\dot\chi
+{1\over2}m^2 \phi\,\chi\bigr),
\label{drho}\\
&&q=\ell_0\,{k\over a}\dot\phi\,\chi\,,
\label{dflux}\\
&&p\,\pi_{L}=\ell_0
\bigl(\dot\phi \,\dot\chi
-{1\over2}m^2 \phi\,\chi\bigr),
\label{dpress}
\end{eqnarray}
in the long wavelength limit.
A notable fact is that these expressions are exactly those one would obtain
for a 4-dimensional scalar field with mass-squared $M^2_{\rm eff}=m^2/2$,
with a simple rescaling $\sqrt{\ell_0}\phi\to\varphi$,
if we neglect the metric perturbation in them,
just as in the case of the homogeneous background.
Thus, provided that we can justify
the neglect of the metric perturbation in these expressions,
the description in terms of the effective
4-dimensional theory with mass-squared $m^2/2$ turns out to be
valid even for inhomogeneous field configurations, at least for
those inhomogeneities whose scales are larger than the Hubble horizon
scale and when the low energy approximation is valid.

Now, let us justify our neglect of the metric perturbation in the
matter variables.
The gauge transformation of the metric perturbation
$h_{\mu\nu}$ induced by an infinitesimal coordination transformation
$x^\mu\to \bar x^\mu=x^\mu+\xi^\mu$ is given by
\begin{eqnarray}
\bar h_{\mu\nu}=h_{\mu\nu}-\xi_{\mu|\nu}-\xi_{\nu|\mu}\,,
\end{eqnarray}
where the vertical bar $|$ denotes the covariant differentiation
with respect to the 4-metric $\gamma_{\mu\nu}$.
Let us assume that a certain `geometrical gauge' is chosen when
we solve the effective 4-dimensional Einstein equations. Here a
`geometrical gauge' is a gauge for which the gauge condition
is given purely in terms of geometrical quantities. An example
is the spatially flat slicing ${\cal R}=H_L+H_T/3=0$ or
the shear-free (Newton) slicing $kB-H_T'=0$.
In such a gauge, since the metric perturbation is due to the scalar
field perturbation $\chi$, we have $h_{\mu\nu}=O(\epsilon\,\phi^2)$,
where $\epsilon$ stands for the amplitude of $\chi$ relative to
$\phi$. Then, for a gauge transformation between two different
geometrical gauges, we have $\xi^\mu=O(\epsilon\,\phi^2)$.
This implies that the change of $\chi$ by such a gauge transformation
is $\bar\chi-\chi=-\phi_{|\mu}\xi^\mu=O(\epsilon\,\phi^3)$.
 Similarly, the change of $\delta T^{\rm eff}_{\mu\nu}$
is of $O(\epsilon\,\phi^4)$ while $\delta T^{\rm eff}_{\mu\nu}$
is of $O(\epsilon\,\phi^2)$.
Hence, to the accuracy of $O(\phi^2)$ that we are interested in,
$\chi$ as well as $\delta T^{\rm eff}_{\mu\nu}$
are gauge-invariant if we restrict our choice of gauge to
a geometrical gauge~\cite{Gen:2002rb}.

The important point to be kept in mind is that the perturbation
equations we derived are valid only in geometrical gauges.
This is because, if we consider
a gauge whose condition involves matter variables, the gauge
transformation from a geometrical gauge to such a
matter-based gauge will give $\xi^\mu=O(\epsilon)$,
and hence $h_{\mu\nu}=O(\epsilon)$. But our equations
can produce $h_{\mu\nu}$ of $O(\epsilon\,\phi^2)$ only.

Nevertheless, we may of course introduce various
gauge-invariant quantities that may be defined in matter-based gauges,
as in the standard cosmological perturbation
theory~\cite{Bardeen:kt,Kodama:bj}.
To repeat again, what we should keep in mind is that
we should evaluate them in a geometrical gauge.
As we see below, the most convenient choice is the flat slicing.
That is, we simply regard our perturbation equations for $\chi$
as those on the flat slicing.

Among various gauge-invariant quantities, a convenient one
for our discussion is the curvature perturbation on the
homogeneous density hypersurface,
first introduced by Wands et al.~\cite{Wands:2000dp},
\begin{eqnarray}
\zeta=H_L+{H_T\over3}+{\rho\,\delta\over3(\rho+p)}
={\cal R}+{\rho\,\delta\over3(\rho+p)}\,.
\label{zetadef}
\end{eqnarray}
The advantage of using this quantity is that
it is known to be conserved in time on superhorizon scales
when the entropy perturbation is negligible, irrespective of
the gravitational theory. Since our $T^{\rm eff}_{\mu\nu}$
is not the real energy-momentum tensor in
the 4-dimensional Einstein theory, using $\zeta$ is preferable.
Then, choosing the flat slicing ${\cal R}=0$,
from Eq.~(\ref{drho}) and the fact that $\rho+p=\ell_0\dot\phi^2$,
we find
\begin{eqnarray}
\zeta={\dot\phi \,\dot\chi
+M_{\rm eff}^2 \phi\,\chi\over 3\dot\phi^2}
\approx
{M_{\rm eff}^2 \phi\,\chi\over 3\dot\phi^2}
\approx -{H\chi\over\dot\phi}\,,
\label{zetasol}
\end{eqnarray}
where we have used the slow-roll equation,
$3H\dot\phi\approx-M_{\rm eff}^2\phi$.
This is the same as the one we would obtain for the standard
4-dimensional theory.

Alternatively, we may choose the curvature perturbation on the
comoving hypersurface defined by~\cite{Kodama:bj,Sasaki:1986hm}
\begin{eqnarray}
{\cal R}_c={\cal R}-{H\chi\over\dot\phi}\,.
\label{calRc}
\end{eqnarray}
Provided that $T^{\rm eff}_{\mu\nu}$ may be regarded as the actual
4-dimensional energy-momentum tensor,
which is indeed appropriate~\cite{Langlois:2000iu},
${\cal R}_c$ is also known to be conversed and equal to $\zeta$
with good accuracy on superhorizon scales.
Again, simply choosing the flat slicing ${\cal R}=0$,
we see that the result is in agreement with Eq.~(\ref{zetasol})
within the accuracy of our interest.

Thus, we conclude that all the predictions we obtain
in this bulk inflaton model are indistinguishable from the
case of the 4-dimensional inflaton model with mass
$M^2_{\rm eff}=m^2/2$,
including the perturbation, in the limit $H^2\ell^2\ll1$.
In particular, one would obtain
the standard, almost scale-invariant spectrum for the large
scale cosmological perturbations.
In this model, possible signatures of the braneworld appears
at most at $O(m^4\ell^2)$, or at second order in the slow-roll
parameter $m^2/H^2$.

\section{Conclusion}

We investigated superhorizon scale cosmological perturbations
in the brane inflation model in which slow-roll inflation on the brane
is induced by the dynamics of a scalar field $\phi$ living in the bulk.
We took the geometrical approach in which the projected Weyl tensor
$E_{\mu\nu}$ describes the gravitational effect on the brane
from the bulk~\cite{Shiromizu:1999wj}.

First we derived the evolution equations for $E_{\mu\nu}$ in the
bulk by assuming a tachionic potential for $\phi$ and
and focusing on the dynamics near the top of the potential
with negative mass-squared $m^2<0$.
We applied them to the case of spatially homogeneous and isotropic
background. We found two different integral forms for $E_{\mu\nu}$,
one with respect to time $t$ and the other to
the extra-dimensional coordinate $z$. By taking the brane limit of the
 $t$-integral form of $E_{\mu\nu}$,
we recovered the results that had been previously obtained
without solving the bulk~\cite{Himemoto:2000nd}.
Namely, in the low energy limit $H^2\ell^2$,
by a rescaling of $\phi$ to an effective 4-dimensional field,
the dynamics on the brane is indistinguishable from
the standard 4-dimensional inflaton model with the
same tachionic potential but multiplied by $1/2$.
In addition, the $z$-integral form of $E_{\mu\nu}$, which
may be useful for analyzing the bulk geometry, is found to give
a new expression of $E_{\mu\nu}$ on the brane that depends
only on the field configuration on the $t=$constant hypersurface.

Then we considered the cosmological perturbations on superhorizon scales.
We found that the effective theory on the brane is still
the same as the case of the homogeneous and isotropic background.
In particular, an anticipated non-trivial contribution from the
spatially anisotropic part of $E_{\mu\nu}$ was found to vanish
at first order in the low energy approximation, i.e., at $O(H^2\ell^2)$.
This implies that possible braneworld signatures may appear
only at $O(H^4\ell^4)$ or higher, or second order
in terms of the slow-roll parameter $m^2/H^2$.
Very recently, based on their low energy expansion
method~\cite{KannoSoda}, and by appealing to the AdS/CFT correspondence,
Kanno and Soda obtained a low energy effective action for
the dilatonic braneworld~\cite{Kanno:03}.
Our result that braneworld signatures may appear only at $O(H^4\ell^4)$
is completely consistent with their recent result~\cite{Kanno:03}.

To predict signatures specific to the braneworld scenario, we thus
have to investigate the effects of $O(H^4\ell^4)$.
Also, it will be interesting to consider the other, high energy limit
$H^2\ell^2\gg1$ in the bulk inflaton model.
Further, it is certainly of interest to explore the geometry in the bulk
in more details, particularly the structure near the future
Cauchy horizon of AdS$_5$.
We hope to come back to these issues in the near future.

\section*{Acknowledgments}
We would like to thank all the participants of
the YITP workshop YITP-W-02-19
on ``Extra dimensions and Braneworld ", held in January 2003,
for valuable discussions.
This work was supported in part by Monbukagaku-sho Grant-in-Aid
for Scientific Research Nos.~12640269 and 14102004.

\appendix

\section{Projected Energy Momentum Tensor}
\label{ProjTab}

In this Appendix, we write down all the projected components
of the energy-momentum tensor that appear in the
effective 4-dimensional Einstein equations
on the background metric given by Eq.~(\ref{bgmetric}).
Specifically, we consider the quadratic part of the energy-momentum tensor,
\begin{eqnarray}
T^{(2)}_{ab}
=\partial_{a} \phi \, \partial_{b}\phi
-g_{ab}
\Bigl(\frac{1}{2}g^{cd}\partial_{c}\phi \,
\partial_{d}\phi +\frac{1}{2}m^2\phi^2 \Bigr) ,
\end{eqnarray}
and write down the components for the spatially homogeneous and inhomogeneous
cases separately.

\subsection{Spatially homogeneous background}

When $\phi$ is spatially homogeneous, we have
\begin{eqnarray}
&&T^{(2)}_{ab}q^{a}_{t} q^{b}_{t}=\frac{1}{2}\Bigl(
\phi^{'2} + \dot{\phi}^2\Bigr) +\frac{1}{2}m^2b^2\phi^2,
\nonumber\\
&&T^{(2)}_{ab}q^{a}_{t} q^{b}_{i}=0,
\nonumber\\
&&T^{(2)}_{ab}q^{a}_{i} q^{b}_{j}= -a^2\Bigl[
\frac{1}{2}\bigl(\phi^{'2}-\dot{\phi}^2 \bigr)
+\frac{1}{2}m^2b^2\phi^2\Bigr]\delta_{ij},
\nonumber\\
&&T^{(2)}_{ab}q^{ab}=-\frac{1}{b^2}\Bigl(2\phi^{'2}-\dot{\phi}^2 \Bigr)
 -2 m^2 \, \phi^2 ,
\nonumber\\
&&T^{(2)}_{ab}q^a_{t} n^b = \frac{\phi^{'} \,\dot{\phi}}{b} ,
\nonumber\\
&&T^{(2)}_{ab}q^a_{i} n^b = 0.
\nonumber\\
&&T^{(2)}_{ab}n^a n^b=\frac{1}{2b^2}\Bigl(\phi^{'2}+\dot{\phi}^2 \Bigr)
-\frac{1}{2}m^2\phi^2,
\nonumber\\
&&T^{(2)}=T^{(2)\,a}{}_a
=-\frac{3}{2b^2}\Bigl(\phi^{'2}-\dot{\phi}^2\Bigr)
-\frac{5}{2}m^2\,\phi^2,
\end{eqnarray}
where the dot ($\dot{~}$) denotes $\partial_t$ and the prime (${}'$)
denotes $\partial_z$.

The 4-dimensional projection of the bulk energy-momentum tensor is
\begin{eqnarray}
T^{\rm(b,2)}_{ab} = \frac{2}{3}
\Bigl[T^{(2)}_{cd} \, q^c_{a} q^d_{b}
 +\bigl(T^{(2)}_{nn}-\frac{1}{4}T^{(2)} \bigr)q_{ab} \Bigr].
\end{eqnarray}
The components of this tensor are given by
\begin{eqnarray}
&&T^{\rm(b,2)}_{tt} = \frac{1}{4}
\Bigl(-\phi^{'2} + \dot{\phi}^2 + m^2b^2 \phi^{2}\Bigr),
\nonumber\\
&&T^{\rm(b,2)}_{ti} = 0,
\nonumber\\
&&T^{\rm(b,2)}_{ij} = \frac{a^2}{12}
\Bigl(3\phi^{'2} +5\dot{\phi}^2 -3  m^2b^2 \phi^{2}\Bigr)
\delta_{ij}.
\nonumber\\
&&T^{\rm(b,2)}=T^{{\rm(b,2)}\,a}{}_a
=\frac{1}{b^2}
\Bigl(\phi^{'2} +\dot{\phi}^2 -m^2 b^2 \phi^2 \Bigr).
\end{eqnarray}

\subsection{Perturbation}
Here we consider the case when $\phi$ has a spatially inhomogeneous
perturbation $\delta\phi$.
The linear perturbation of $T^{(2)}_{ab}$ is
\begin{eqnarray}
\delta T^{(2)}_{ab}
=2\partial_{(a} \phi \,\partial_{b)}\delta\phi
-g_{ab}
\Bigl(g^{cd}\partial_{c}\phi \partial_{d}\delta \phi
+m^2\phi \,\delta \phi \Bigr).
\end{eqnarray}
The components are
\begin{eqnarray}
&&\delta T^{(2)}_{ab}q^{a}_{t} q^{b}_{t}=\Bigl(
\phi^{'}\, {\delta\phi}^{'}
+ \dot{\phi}\,\dot{\delta\phi}\Bigr) +m^2b^2\phi\,\delta\phi,
\nonumber\\
&&\delta T^{(2)}_{ab}q^{a}_{t} q^{b}_{i}= \dot{\phi}\,
\partial_i\delta{\phi},
\nonumber\\
&&\delta T^{(2)}_{ab}q^{a}_{i} q^{b}_{j}= -a^2\Bigl[
\bigl(\phi^{'}{\delta\phi}^{'}-\dot{\phi}\,\dot{\delta\phi} \bigr)
+m^2b^2\phi \,\delta\phi\Bigr] \delta_{ij},
\nonumber\\
&&\delta T^{(2)}_{ab}q^{ab}
=-\frac{2}{b^2}\Bigl(2\phi^{'} \delta\phi^{'}
-\dot{\phi} \,  \dot{\delta \phi} \Bigr)-4 m^2 \, \phi \delta \phi .
\nonumber\\
&&\delta T^{(2)}_{ab}q^a_{t} n^b =
\frac{\phi^{'} \dot{\delta\phi}+\dot{\phi}\,{\delta\phi}^{'}}{b},
\nonumber\\
&&\delta T^{(2)}_{ab}q^a_{i} n^b =\frac{\phi^{'}
\partial_i\delta\phi}{b}.
\nonumber\\
&&\delta T^{(2)}_{ab}n^an^b
=\frac{1}{b^2}\Bigl(\phi^{'}\,\delta\phi^{'}
+\dot{\phi}\, \dot{\delta\phi} \Bigr)-m^2\phi\,\delta{\phi}.
\nonumber\\
&&\delta T^{(2)}=-\frac{3}{b^2}\Bigl( \phi^{'} \delta \phi^{'}
-\dot{\phi} \, \dot{\delta \phi}\Bigr)- 5m^2 \phi\,\delta\phi.
\end{eqnarray}

The perturbation of the 4-dimensional projection of the bulk
 energy-momentum tensor is
\begin{eqnarray}
\delta T^{\rm(b,2)}_{ab} = \frac{2}{3}
\Bigl[\delta T^{(2)}_{cd} \, q^c_{a} q^d_{b}
 +\bigl(\delta T^{(2)}_{nn}-\frac{1}{4} \delta \, T^{(2)} \bigr)
q_{ab} \Bigr].
\end{eqnarray}
The components are
\begin{eqnarray}
&&\delta T^{\rm(b,2)}_{tt} = \frac{1}{2}
\Bigl(-\phi^{'}{\delta\phi}^{'} + \dot{\phi}\,\dot{\delta\phi}
+ m^2b^2 \phi\,\delta\phi\Bigr),
\nonumber\\
&&\delta T^{\rm(b,2)}_{ti} =\frac{2}{3}
\dot{\phi}\,\partial_i{\delta \phi},
\nonumber\\
&&\delta T^{\rm(b,2)}_{ij} = \frac{a^2}{6}
\Bigl(3\phi^{'}{\delta \phi}^{'}+5\dot{\phi}\,\dot{\delta\phi}
-3  m^2b^2 \phi \,\delta\phi\Bigr) \delta_{ij} ,
\nonumber\\
&&\delta T^{\rm(b,2)}=\frac{2}{b^2}
\Bigl(\phi^{'}\delta\phi^{'} +\dot{\phi}\,\dot{\delta \phi}
-m^2 b^2 \phi\, \delta \phi  \Bigr).
\end{eqnarray}

\section{Source terms for $\bm{E_{\mu\nu}}$}
\label{Source}

Here, we give explicit expressions of the terms that contribute
to the source term $S_{\mu\nu}$ in Eq.~(\ref{2ndEeq}) for $E_{\mu\nu}$
in the bulk.
Again, we treat the spatially homogeneous background
and the perturbation separately.

The tensors that appear in the source term are
\begin{eqnarray}
P_{\mu\nu}[T^{(2)}_{ab}]
&=&
-\frac{8}{3}\frac{b'}{b^2}\Bigl(T^{(2)}_{nn}-\frac{1}{4}T^{(2)}\Bigr)
q_{\mu\nu}
-\frac{2}{3}\frac{b'}{b^2}T^{(2)}_{\mu\nu}
-\frac{1}{3}\frac{b'}{b^2}T^{(2)\alpha}_{\alpha}q_{\mu\nu}
-\frac{2}{3}{\pounds}_{n}\Bigl(T^{(2)}_{nn}
-\frac{1}{4}T^{(2)}\Bigr)q_{\mu\nu}
\nonumber\\
&&+\frac{2}{3}{\pounds_{n}}\Bigl(T^{(2)}_{\mu\nu}\Bigr)
-\frac{1}{3}q_{\mu\nu}{\pounds_{n}}T^{(2)\alpha}_{\alpha}
-\frac{2}{3}D^{\alpha}\Bigl(T^{(2)}_{n[\alpha}q_{(\mu]\nu)}\Bigr)
-D_{(\mu}T^{(2)}_{\nu)n}\,,
\nonumber\\
Q_{\mu\nu\alpha}[T^{(2)}_{ab}]
&=&\frac{2}{3}
{\pounds_{n}}\Bigl(T^{(2)}_{n[\mu}q_{(\nu]\alpha)}\Bigr)
-\frac{2}{3}D_{\mu}
\Bigl(q_{(\nu]\alpha)}\Bigl(T^{(2)}_{nn}-\frac{1}{2}T^{(2)}\Bigr) \Bigr)
-D_{[\mu}T^{(2)}_{(\nu]\alpha)}\,,
\nonumber\\
\Sigma_{\mu}[T^{(2)}_{ab}]
&=&D^{\alpha}T^{(b,2)}_{\mu\alpha}
+2\frac{b'}{b^2}T^{(2)}_{n a}q^{a}_{\mu}\,,
\end{eqnarray}
where $Q_{\mu\nu\alpha}$ contributes to the source term only through
the form $D^{\alpha}Q_{\alpha (\mu\nu)}$.

\subsection{Spatially homogeneous background}

For the spatially homogeneous background discussed in Sec.~III~C,
the non-vanishing components are
\begin{eqnarray}
P_{tt}[T^{(2)}_{ab}]
&=&-\frac{1}{2}m^2 b \phi\phi'
+\frac{3b'}{2b^2}\phi'^2
+\frac{\phi'\phi''}{2b}
-\frac{\dot{a}\phi'\dot{\phi}}{ab}
-\frac{b'\dot{\phi}^2}{2b^2}
+\frac{\dot{\phi}\dot{\phi}'}{2b}
-\frac{\phi'\ddot{\phi}}{b}\,,
\nonumber\\
D^{\alpha}Q_{\alpha tt}[T^{(2)}_{ab}]
&=&
\frac{\dot{a}}{a}\Bigl(
-\frac{1}{2}m^2\phi\dot{\phi}
+\frac{b'}{b^3}\phi' \dot{\phi}
+\frac{1}{b^2}\phi''\dot{\phi}
-\frac{\dot{a}}{ab^2}\dot{\phi}^2
-\frac{1}{2 b^2}\phi'\dot{\phi'}
-\frac{1}{2b^2}\dot{\phi}\ddot{\phi}
\Bigr)\,,
\nonumber\\
\Sigma_{t}[T^{(2)}_{ab}]
&=&-\frac{1}{2}m^2\phi\dot{\phi}
+2\frac{b'}{b^3}\phi'\dot{\phi}
-2\frac{\dot{a}}{ab^2}\dot{\phi}^2
+\frac{1}{2b^2}\phi'\dot{\phi'}
-\frac{1}{2b^2}\dot{\phi}\ddot{\phi}\,.
\end{eqnarray}
All the other components of $P_{\mu\nu}$,
$D^{\alpha}Q_{\alpha (\mu\nu)}$ and $\Sigma_\mu$
vanish.

\subsection{Perturbation}

We list the explicit expressions for the source terms
under the presence of a perturbation $\delta\phi=\chi(t,z)Y(x^i)$.
As discussed in Sec.~IV, we adopt the expansion in terms of
spatial harmonics $Y(x^i)$.

The relevant coefficients of the harmonic expansion are defined in
Eqs.~(\ref{deltaP}), (\ref{deltaDQ}) and (\ref{Sform}).
They are given by
\begin{eqnarray}
P&=&-\frac{1}{2} m^2b \phi'\chi -\frac{1}{2}m^2 b\phi\chi'
+\frac{3b'}{b^2}\phi'\chi' +\frac{1}{2b}{\phi''\chi'}
-\frac{\dot{a}}{ab}\dot{\phi}\chi'-\frac{1}{b}\ddot{\phi}\chi'
+\frac{1}{2b}\phi'\chi''
\nonumber\\
&&\qquad-\frac{\dot{a}}{ab}\phi' \dot{\chi}
-\frac{b'}{b^2} \dot{\phi}\dot{\chi}
+\frac{1}{2b}\dot{\phi'}\dot{\chi}
+\frac{1}{2b}\dot{\phi}\dot{\chi}'-\frac{1}{b}\phi'\ddot{\chi}
-\frac{k^2}{3a^2b}\phi'\chi\,,
\nonumber\\
P_{1}
&=&-\frac{k}{a}\,
\Bigl(\frac{\dot{a}}{a}\frac{2}{3b}\phi'\chi
-\frac{2b'}{3b^2}\dot{\phi}\chi+\frac{1}{3b}\dot{\phi'}\chi
+\frac{1}{3b}\dot{\phi}\chi'-\frac{2}{3b}\phi'\dot{\chi}\Bigr)\,,
\nonumber \\
P_{2}
&=&-\frac{2}{3b}\,\frac{k^2}{a^2}\phi'\chi\,.
\\
\cr\cr
Q&=&-\frac{1}{2}\frac{\dot{a}}{a}m^2\dphi\chi
+\frac{\dot{a}}{a}\frac{b'}{b^3}\dot{\phi}\chi'
-\frac{\dot{a}}{a}\frac{1}{2b^2}\dot{\phi}'\chi'
+\frac{\dot{a}}{a}\frac{1}{b^2}\dot{\phi}\chi''
-\frac{1}{2}\frac{\dot{a}}{a}m^2\phi\dot{\chi}
\nonumber\\
&&\quad
+\frac{b'}{b^3}\frac{\dot{a}}{a}\phi'\dot{\chi}
+\frac{\dot{a}}{a}\frac{1}{b^2}\phi''\dot{\chi}
-\frac{\dot{a}^2}{a^2}\frac{2}{b^2}\dot{\phi}\dot{\chi}
-\frac{\dot{a}}{a}\frac{1}{2b^2}\ddot{\phi}\dot{\chi}
-\frac{\dot{a}}{a}\frac{1}{2b^2}\phi'\dot{\chi}'
-\frac{\dot{a}}{2a}\frac{1}{b^2}\dot{\phi}\ddot{\chi}
\nonumber\\
&&\quad
-\frac{k^2}{a^2}\,\Bigl[
\frac{m^2}{6}\phi\chi
-\frac{b'}{3b^3}\phi'\chi
-\frac{1}{3b^2}\phi''\chi
+\frac{\dot{a}}{a}\frac{2}{3b^2}\dphi\chi
+\frac{1}{3b^2}\ddot{\phi}\chi
+\frac{1}{6b^2}\phi'\chi'
-\frac{1}{6b^2}\dot{\phi}\dot{\chi}
\Bigr]\,,
\nonumber\\
Q_{1}&=&- \frac{k}{a}\,
\Bigl[
-\frac{m^2}{6}\frac{\dot{a}}{a}\phi\chi
+\frac{\dot{a}}{a}\frac{b'}{3 b^3}\phi' \chi
+\frac{\dot{a}}{a}\frac{1}{3b^2}\phi''\chi
+\frac{m^2}{6}\dot{\phi}\chi
-\frac{\dot{a}^2}{a^2}\frac{1}{2b^2}\dot{\phi}\chi
+\frac{\ddot{a}}{a}\frac{1}{6b^2}\dphi\chi
-\frac{b'}{6b^3}\dot{\phi'}\chi
\nonumber\\
&&\quad
-\frac{1}{6b^2}\dot{\phi''}\chi
+\frac{1}{6b^2}\frac{\dot{a}}{a}\ddot{\phi}\chi
+\frac{1}{6b^2}\dot{\ddot{\phi}}\chi
-\frac{\dot{a}}{a}\frac{1}{6b^2}\phi'\chi'
-\frac{b'}{6b^3}\dphi\chi'
+\frac{1}{6b^2}\dot{\phi}'\chi'
-\frac{1}{6b^2}\dot{\phi}\chi''
\nonumber\\
&&\quad
+\frac{1}{6}m^2\phi\dot{\chi}
-\frac{b'}{3b^3}\phi'\dot{\chi}
-\frac{1}{3b^2}\phi''\dot{\chi}
+\frac{\dot{a}}{a}\frac{1}{3b^2}\dot{\phi}\dot{\chi}
+\frac{1}{6b^2}\ddot{\phi}\dot{\chi}
+\frac{1}{6b^2}\phi'\dot{\chi}'
+\frac{k^2}{a^2}\frac{1}{6b^2}\dot{\phi}\chi
\Bigr]\,,
\nonumber\\
Q_{2}&=&\frac{k^2}{a^2}\,
\Bigl[
\frac{1}{6}m^2\phi\chi
-\frac{b'}{3b^3}\phi'\chi
+\frac{1}{6b^2}\phi'\chi'
-\frac{1}{3b^2}\phi''\chi
+\frac{\dot{a}}{a}\frac{1}{3b^2}\dot{\phi}\chi
-\frac{1}{6b^2}\dot{\phi}\dot{\chi}
-\frac{1}{3b^2}\ddot{\phi}\chi
\Bigr]\,.
\\
\cr\cr
\Sigma&=&-\frac{1}{2}m^2\dphi\chi-\frac{1}{2}m^2\phi\dot{\chi}
+\frac{2b'}{b^3}\dphi\chi'
+\frac{1}{2b^2}\dphi'\chi'+\frac{2b'}{b^3}\phi'\dot{\chi}
-\frac{4\dot{a}}{a}\frac{1}{b^2}\dphi\dot{\chi}
-\frac{1}{2b^2}\ddphi\dot{\chi}
\nonumber\\
&&\qquad+\frac{1}{2b^2}\phi'\dot{\chi'}
-\frac{1}{2b^2}\dot{\phi}\ddot{\chi}
-\frac{2}{3b^2}\frac{k^2}{a^2}\dphi\chi\,,
\nonumber\\
\Sigma_{1}&=&-\frac{k}{a}\,\Bigl[-\frac{1}{2}m^2\phi\chi
+\frac{2b'}{b^3}\phi'\chi-\frac{2\dot{a}}{a}\frac{1}{b^2}\dphi\chi
-\frac{2}{3b^2}\ddphi\chi+\frac{1}{2b^2}\phi'\chi'
+\frac{1}{6b^2}\dphi\dot{\chi} \Bigr]\,.
\end{eqnarray}

\section{Green functions for $E_{\mu\nu}$}
\label{Greenfcn}

In this Appendix, we analyze the late time behavior of
the retarded Green functions for Eq.~(\ref{Ettreq}) for $E_{tt}$
and Eq.~(\ref{E2longeq}) for the spatially anisotropic part
$E_2Y_{ij}$. We consider the long wavelength limit.
Hence Eq.~(\ref{Ettreq}) is the same as that for the spatially
homogeneous case, Eq.~(\ref{Etthomoeq}).
Since the difference between these two Green functions is
rather trivial, we mainly analyze the Green function for $E_{tt}$.
The Green function for $E_2$ is given in the last subsection.
Our analysis here is parallel to
the one given in~\cite{Himemoto:2001hu}.

\subsection{Mode functions}

We construct the Green function from a superposition of mode functions
that satisfy the source-free equation,
\begin{eqnarray}
\Bigl[b(z)\partial_{z} \Bigl(\frac{1}{b(z)}\partial_{z}\Bigr)
 - \partial_{t}^2 -7 H \partial_{t}
-12 H^2\Bigr] \Psi(z,t)= 0\,.
\label{homoeq}
\end{eqnarray}
Putting $\Psi=u_q(z)\psi_q(t)$,
we have
\begin{eqnarray}
\Bigl[b(z)\partial_{z}\Bigl(\frac{1}{b(z)}\partial_{z} \Bigr)
+(n^2-2)H^2  \Bigr] u_{q}(z)=0\,,
 \label{uzeq}\\
\Bigl[\partial_{t}^2 +7 H \partial_{t} +(n^2+10)H^2 \Bigr]\psi_{q}(t)=0\,,
\label{psieq},
\end{eqnarray}
where we have put $n^2=q^2+9/4$ for later convenience.

Equation~(\ref{uzeq}) can be rewritten in the
standard Schr\"{o}dinger from by setting
$v_{q}(z)=b^{-1/2}(z)u_{q}(z)$,
\begin{eqnarray}
\bigl[-\frac{d^2}{dz^2}+ H^2 V(z)\bigr]v_{q}(z)
=n^2 H^2 v_{q}(z)\,,
\label{Schform}
\end{eqnarray}
where
\begin{eqnarray}
V(z)= \frac{9}{4}-\frac{1}{4\sinh^2\bigl[H(|z|+z_0)\bigr]}
+H \coth\bigl[H(|z|+z_0)\bigr] \delta(z)\,.
\label{antivolcano}
\end{eqnarray}
The potential has a positive delta-function singularity at $z=0$,
i.e., it has an `anti-volcano' form. It then follows that
there exists no bound-state mode~\cite{Gen:2000nu},
unlike the case of the bulk scalar field itself.
As the potential approaches $9/4$ at $|z|\to\infty$,
$n^2>9/4$ and the spectrum is continuous.
Thus, we may label the mode functions
in terms of $q=\pm\sqrt{n^2-9/4}$ as $\Psi_q(t,z)=u_q(z)\psi_q(t)$
with $-\infty<q<\infty$.

There are two independent mode functions belonging to each eigenvalue $q$.
For a given $q$, we set
\begin{eqnarray}
\psi_{q} (t) = \frac{1}{\sqrt{2 \pi}}e^{-i q H t}e^{-\frac{7}{2}H t}\,,
 \label{psiq}
\end{eqnarray}
which satisfies the ortho-normality and completeness,
\begin{eqnarray}
\int_{-\infty}^{\infty} dt\, H e^{7 H t} \psi_{q}(t)\psi^{\ast}_{q^{'}}(t)
= \delta(q-q^{'}) \,,
\quad
\int_{-\infty}^{\infty} dq\, \psi_{q}(t)\psi^{\ast}_{q}(t')=
\frac{\delta(t-t')}{H e^{7 H t}}\,.
 \label{orthcomp}
\end{eqnarray}
With this choice of $\psi_q(t)$,
a convenient choice for the two independent solutions
is to require the outgoing-wave and
ingoing-wave boundary conditions at $|z|=\infty$.
Specifically,
\begin{eqnarray}
&&u_{q}^{({\rm out})}(z)
= \frac{1}{\sqrt{\pi}}\frac{\Gamma(1-iq)}{\Gamma(1/2-iq)}
Q_{-1/2-iq}(\xi)
=2^{-\frac{1}{2}+iq} \xi^{-\frac{1}{2}+iq}
  {}_{2} F_{1}
\bigl[\frac{\frac{1}{2}-iq}{2},\frac{\frac{3}{2}-iq}{2};
1-iq,\frac{1}{\xi^2}\bigr],
\label{uqout}
 \\
&& u_{q}^{({\rm in})}(z)
= \frac{1}{\sqrt{\pi}}\frac{\Gamma(1+iq)}{\Gamma(1/2+iq)}
Q_{-1/2+iq}(\xi)
=u_{-q}^{({\rm out})}(z),
\label{uqin}
\end{eqnarray}
where $Q_{\nu}(\xi)$ is the Legendre function of the second kind
and ${}_{2}F_{1}[\alpha,\beta;\gamma,y]$ denotes the
hypergeometric function~\cite{Bateman:1969ku}, and
$\xi \equiv \cosh \bigl[H\bigl(|z|+z_{0} \bigr) \bigr]$.
Since the regions $z>0$ and $z<0$ are identical, we focus on
the region $z>0$ in what follows.

\subsection{Construction of the Green function}

We now construct the retarded
Green function introduced in Eq. (\ref{GretEtt})
from a superposition over the modes,
\begin{eqnarray}
G\bigl(t,z;t',z' \bigr)
= \int_{-\infty}^{\infty}dq\,  G_{q}(z;z')\,
\psi_{q}(t)\psi_{-q}(t')a^7(t')\,,
\label{Gsup}
\end{eqnarray}
where $G_{q}(z,z')$ satisfies
\begin{eqnarray}
\bigl[b(z)\partial_{z}\Bigl(\frac{1}{b(z)}\partial_{z}
\Bigr)+(n^2 - 2)H^2 \bigr]
G_{q}(z;z')=-\frac{H^9 l}{\sinh[H(z+z_{0})]}\delta(z-z')
\label{Gqeq}.
\end{eqnarray}
To construct $G_q(z,z')$, we need to require the $Z_{2}$-symmetry, or
the Neumann boundary condition on the brane, in addition to
the outgoing-wave boundary condition at (conformal) infinity.
The $Z_2$-symmetric mode function is given by
\begin{eqnarray}
u^{(Z_{2})}(z)=u^{({\rm out})}(z)-\gamma_{q} u^{({\rm in})}(z)
\,,
\quad
\gamma_{q}\equiv \frac{\partial_{z} u^{({\rm out})}(z)}
 {\partial_{z} u^{({\rm in})}(z)} \Bigg|_{z=0} \,.
\label{uZ2}
\end{eqnarray}
Then we have
\begin{eqnarray}
G_{q}(z;z')=\frac{1}{W_{q}}
\bigl[u_{q}^{({\rm out})}(z') u_{q}^{(Z_2)}(z)\theta(z'-z)
+u_{q}^{({\rm out})}(z) u_{q}^{(Z_2)}(z')\theta(z-z')
\bigr]\,,
\label{Gqform}
\end{eqnarray}
where $W_{q}$ is the Wronskian defined by
\begin{eqnarray}
&&W_{q} \equiv \frac{\sinh\bigl[H\bigl(z+z_{0}\bigr) \bigr]}{H^9 l}
\bigl[\partial_{z}u^{(Z_2)}(z)u_{q}^{({\rm out})}(z)
 -u_{q}^{(Z_2)}(z)\partial_{z}u_{q}^{({\rm out})}(z)\bigr]
=\frac{iq}{Hl} \frac{\gamma_{q}}{H^7}\,.
\label{Wrons}
\end{eqnarray}

The behavior of the Green function at late times can be studied
by analyzing poles of $G_q$, or equivalently zeros of the Wronskian,
in the complex $q$-plane~\cite{Himemoto:2001hu}.
For this purpose, here, we give the $z$-derivative of
$u^{\rm(out)}$ explicity,
\begin{eqnarray}
\partial_{z} u^{({\rm out})}(z)
&=& \bigl(-\frac{1}{2} +i q \bigr)2^{-\frac{1}{2}+iq}\xi^{-\frac{3}{2}+iq}
\sinh \bigl[H\bigl(z+z_{0} \bigr) \bigr]
\nonumber\\
&&\quad
\times\Biggl(
 {}_{2}F_{1}
\left[\frac{\frac{1}{2}-iq}{2},\frac{\frac{3}{2}-iq}{2};1-iq,\frac{1}{\xi^2}
\right]
\nonumber\\
&&\qquad\quad
+ \xi^{-2} \frac{\frac{1}{2}(\frac{3}{2}-iq)}{1-iq}
{}_{2} F_{1}
\left[\frac{\frac{5}{2}-iq}{2},\frac{\frac{7}{2}-iq}{2};2-iq,\frac{1}{\xi^2}
\right]
 \Biggr).
\label{dzuout}
\end{eqnarray}
{}From this, expressions for $\partial_{z} u^{({\rm in})}(z)$ and
$\partial_z u^{(Z_2)}$ follow immediately.

Hereafter, we consider the late time behavior of the Green function
in the cases $Hl\ll 1$ and $Hl\gg 1$ separately.

\vspace{3mm}
\noindent
{\bf 1. The case $\bm{Hl\ll 1}$}
\vspace{3mm}

{}From Eq.~(\ref{dzuout}),
\begin{eqnarray}
\partial_{z} u^{({\rm out})}(0+)
\approx 2^{-\frac{1}{2}+iq}H \sinh\bigl(H z_{0}\bigr) (-\frac{1}{2}+iq)
\xi_{0}^{-3/2+i q}
\frac{\Gamma(1-iq)}{\Gamma(\frac{\frac{3}{2}-iq}{2})
\Gamma(\frac{\frac{5}{2}-iq}{2})(H l)^2 } \,,
\label{dzulimit}
\end{eqnarray}
and similarly for $\partial_{z} u^{({\rm in})}(0+)$.
Thus, in the limit $H\ell\ll1$,
\begin{eqnarray}
\gamma_{q}= \frac{-1/2+iq}{-1/2-iq}\,
\frac{\Gamma(1-iq)\Gamma(\frac{3}{2}+iq)}
 {\Gamma(1+iq)\Gamma(\frac{3}{2}-iq)}\,,
\label{gammalimit}
\end{eqnarray}
and the Green function $G_q$ becomes
\begin{eqnarray}
G_{q}(z;z')&=& \bigl(H l\bigr) H^7
\bigl[i\pi \tanh(q \pi)P_{-\frac{1}{2}+iq}(\xi)P_{-\frac{1}{2}+iq}(\xi^{'})
\nonumber\\
&&\quad
+\bigl( P_{-\frac{1}{2}+i q}(\xi)Q_{-\frac{1}{2}+i q}(\xi^{'})
\theta(z'-z)
+Q_{-\frac{1}{2}+i q}(\xi)P_{-\frac{1}{2}+i q}(\xi^{'})
\theta(z-z') \bigr)\bigr].
\end{eqnarray}
The total Green function is given by Eq.~(\ref{Gsup}), i.e., by
integral with respect to $q$ along the whole real axis
on the complex $q$-plane. The retarded condition is guaranteed
if there is no pole in the upper half of complex $q$-plane,
which is in fact the case.

For $t>t'$, the integral (\ref{Gsup}) is equivalent to
the contour integral over the whole region of the lower half
complex $q$-plane, or the sum of the residues of these poles.
In the limit $H\ell\ll1$, we find that
$G_q$ has poles only at
\begin{eqnarray}
q= -i \frac{2s+1}{2}\quad(s=0,1,2,\cdots).
\end{eqnarray}
Hence at $t>t'$, we may express the
retarded Green function as
\begin{eqnarray}
G(t,z;t',z')=(Hl) \theta(t-t')
\sum_{n=0}^{\infty}P_{n}(\xi)P_{n}(\xi')
a^{-(n+4)}(t)a^{n+4}(t'),
\end{eqnarray}
where $P_{n}(z)$ is Legendre polynomial of order $n$.
In particular, after a sufficient lapse of time,
we may approximate the Green function
the $n=0$ term, i.e., by taking account of only the pole with
the smallest imaginary part $q=-i/2$.
This gives
\begin{eqnarray}
G\bigl(t,z;t',z' \bigr)
 \approx (H l)\,\theta(t-t')\, a^{-4}(t)a^{4}(t')\,.
 \label{Glatetime}
\end{eqnarray}
Thus, $G \propto a^{-4}$ at sufficiently late times.
This gives rise to the radiation-like behavior of $E_{tt}$,
in accordance with our expectation.

\vspace{3mm}
\noindent
{\bf 2. The case \bm{$Hl\gg 1$}}
\vspace{3mm}

In this case, we have
\begin{eqnarray}
\gamma_{q}=\frac{-1/2+i q}{-1/2-i q}\, 2^{2iq}\, \xi_{0}^{2iq}\,,
\label{gammalarge}
\end{eqnarray}
and the structure of $G_{q}(z,z')$ is more complicated than
the case $H\ell\ll1$. Nevertheless, it is possible to locate
the poles of $G_q$. We find the poles at
\begin{eqnarray}
q=-\frac{i}{2}\,,\quad
-is\quad (s=1,2,3,\cdots).
\end{eqnarray}
Thus at sufficiently late times,
the Green function behaves as
\begin{eqnarray}
G\bigl(t,z;t',z' \bigr)\approx
 \theta(t-t')\,a^{-4}(t)a^{4}(t')\, .
\label{Glarge}
\end{eqnarray}
Apart from the absence of the factor $Hl$,
this is the same as that in the case $H\ell\ll1$, Eq.~(\ref{Glatetime}).
Hence the radiation-like behavior is realized again.
Of course, this is also expected, because the transverse-traceless
nature of the source-free $E_{\mu\nu}$ is independent of the
energy scale.

\subsection{Green function for $\bm{E_2}$}

The Green function $G_{2}(t,z;t',z')$ for $E_2$
satisfies
\begin{eqnarray}
\Bigl[b(z)\partial_{z}\Bigl(\frac{1}{b(z)}\partial_{z}\Bigr)
-\partial_{t}^2-3H\partial_{t}-2H^2\Bigr] G_{2}(t,z;t',z')
=-\delta(t-t')\delta(z-z')b(z),
\end{eqnarray}
in the long wavelength limit.
In this case, only the difference from the Green function for $E_{tt}$
appears in the temporal part of the mode functions.
For an eigenvalue $q$, with $q^2=n^2-9/4$ as before,
the $z$-component of the equation is exactly the same
as Eq.~(\ref{uzeq}), while the $t$-component becomes
\begin{eqnarray}
\Bigl[\partial_{t}^2 +3 H \partial_{t} +n^2H^2 \Bigr]\psi_{2,q}(t)=0.
\end{eqnarray}
The solution is related to $\psi_{q}(t)$ for $E_{tt}$
as $\psi_{2,q}(t)=a^{2}(t)\psi_{q}(t)$.
Hence, it follows that
\begin{eqnarray}
G_2(t,z;t',z')=a^2(t)G(t,z;t'z')a^{-2}(t')\,,
\end{eqnarray}
and we can use all the results in the previous subsections.
In particular, in the limit $H\ell\ll1$,
the Green function at $t>t'$ is expressed as
\begin{eqnarray}
G_{2}(t,z;t',z')=(Hl)\theta(t-t')
 \sum_{n=0}^{\infty}P_{n}(\xi)P_{n}(\xi')
a^{-(n+2)}(t)a^{n+2}(t').
\label{G2late}
\end{eqnarray}
Also, at sufficiently late times, we have
$G_{2}\propto a^{-2}$, irrespective to $Hl$.


\begin{thebibliography}{99}


\bibitem{braneworld}
K.~Akama,
Lect.\ Notes Phys.\  {\bf 176}, 267 (1982)
[arXiv:hep-th/0001113].
\\
V.~A.~Rubakov and M.~E.~Shaposhnikov,
Phys.\ Lett.\ B {\bf 125}, 139 (1983).
\\
G.~W.~Gibbons and D.~L.~Wiltshire,
Nucl.\ Phys.\ B {\bf 287}, 717 (1987)
[arXiv:hep-th/0109093].
\\
P.~Horava and E.~Witten,
Nucl.\ Phys.\ B {\bf 460}, 506 (1996)
[arXiv:hep-th/9510209].
\\
P.~Horava and E.~Witten,
Nucl.\ Phys.\ B {\bf 475}, 94 (1996)
[arXiv:hep-th/9603142].
\\
N.~Arkani-Hamed, S.~Dimopoulos and G.~R.~Dvali,
Phys.\ Lett.\ B {\bf 429}, 263 (1998)
[arXiv:hep-ph/9803315].
\\
I.~Antoniadis, N.~Arkani-Hamed, S.~Dimopoulos and G.~R.~Dvali,
Phys.\ Lett.\ B {\bf 436}, 257 (1998)
[arXiv:hep-ph/9804398].
\\
L.~Randall and R.~Sundrum,
Phys.\ Rev.\ Lett.\  {\bf 83}, 3370 (1999)
[arXiv:hep-ph/9905221].

\bibitem{Randall:1999vf}
L.~Randall and R.~Sundrum,
Phys.\ Rev.\ Lett.\ {\bf 83}, 4690 (1999)
[arXiv:hep-th/9906064].

\bibitem{bwcosreview}
For a nice review on braneworld cosmology, see D.~Langlois,
to be published in Prog. Theor. Phys. Suppl. No.~148
[arXiv:hep-th/0209261].

\bibitem{Garriga:1999yh}
J.~Garriga and T.~Tanaka,
Phys.\ Rev.\ Lett.\  {\bf 84}, 2778 (2000)
[arXiv:hep-th/9911055].

\bibitem{KT}
For a review, see, e.g.,
A.~R.~Liddle and D.~H.~Lyth,
 {\it Cosmological Inflation and Large Scale Structure}
(Cambridge University Press, Cambridge, 2000).

\bibitem{braneinf}
R.~Maartens, D.~Wands, B.~A.~Bassett and I.~Heard,
Phys.\ Rev.\ D {\bf 62}, 041301 (2000)
[arXiv:hep-ph/9912464].
\\
R.~Dvali and S.~H.~Tye,
Phys.\ Lett.\ B {\bf 450}, 72 (1999)
[arXiv:hep-ph/9812483].
\\
S.~W.~Hawking, T.~Hertog and H.~S.~Reall,
Phys.\ Rev.\ D {\bf 63}, 083504 (2001)
[arXiv:hep-th/0010232].
\\
N.~Kaloper,
Phys.\ Rev.\ D {\bf 60}, 123506 (1999)
[arXiv:hep-th/9905210].
\\
H.~B.~Kim and H.~D.~Kim,
Phys.\ Rev.\ D {\bf 61}, 064003 (2000)
[arXiv:hep-th/9909053].
\\
H.~A.~Chamblin and H.~S.~Reall,
Nucl.\ Phys.\ B {\bf 562}, 133 (1999)
[arXiv:hep-th/9903225].
\\
S.~Nojiri and S.~D.~Odintsov,
Phys.\ Lett.\ B {\bf 484}, 119 (2000)
[arXiv:hep-th/0004097].
\\
G.~Huey and J.~E.~Lidsey,
Phys.\ Lett.\ B {\bf 514}, 217 (2001)
[arXiv:astro-ph/0104006].
\\
A.~Lukas and D.~Skinner,
JHEP {\bf 0109}, 020 (2001)
[arXiv:hep-th/0106190].
\\
M.~C.~Bento, O.~Bertolami and A.~A.~Sen,
arXiv:hep-th/0208124.
\\
A.~R.~Liddle and L.~A.~Urena-Lopez,
arXiv:astro-ph/0302054.

\bibitem{Garriga:1999bq}
J.~Garriga and M.~Sasaki,
Phys.\ Rev.\ D {\bf 62}, 043523 (2000)
[arXiv:hep-th/9912118].

\bibitem{bwcreation}
K.~Koyama and J.~Soda,
Phys.\ Lett.\ B {\bf 483}, 432 (2000)
[arXiv:gr-qc/0001033].
\\
S.~Kanno, M.~Sasaki and J.~Soda,
arXiv:hep-th/0210250.
\\
R.~Brandenberger, G.~Geshnizjani and S.~Watson,
arXiv:hep-th/0302222.



\bibitem{Himemoto:2000nd}
Y.~Himemoto and M.~Sasaki,
Phys.\ Rev.\ D {\bf 63}, 044015 (2001)
[arXiv:gr-qc/0010035];

\bibitem{bwqfluc}
S.~Kobayashi, K.~Koyama and J.~Soda,
Phys.\ Lett.\ B {\bf 501}, 157 (2001)
[arXiv:hep-th/0009160].
\\
N.~Sago, Y.~Himemoto and M.~Sasaki,
Phys.\ Rev.\ D {\bf 65}, 024014 (2002)
[arXiv:gr-qc/0104033].

\bibitem{LanRod01}
D.~Langlois and M.~Rodriguez-Martinez,
Phys.\ Rev.\ D {\bf 64}, 123507 (2001)
[arXiv:hep-th/0106245].

\bibitem{Himemoto:2001hu}
Y.~Himemoto, T.~Tanaka and M.~Sasaki,
Phys.\ Rev.\ D {\bf 65}, 104020 (2002)
[arXiv:gr-qc/0112027].

\bibitem{bwreheat}
J.~Yokoyama and Y.~Himemoto,
Phys.\ Rev.\ D {\bf 64}, 083511 (2001)
[arXiv:hep-ph/0103115].
\\
Y.~Himemoto and T.~Tanaka,
arXiv:gr-qc/0212114.

\bibitem{KoyTak03}
K.~Koyama and K.~Takahashi,
arXiv:hep-th/0301165.

\bibitem{Shiromizu:1999wj}
T.~Shiromizu, K.~i.~Maeda and M.~Sasaki,
Phys.\ Rev.\ D {\bf 62}, 024012 (2000)
[arXiv:gr-qc/9910076];
\\
M.~Sasaki, T.~Shiromizu and K.~i.~Maeda,
Phys.\ Rev.\ D {\bf 62}, 024008 (2000)
[arXiv:hep-th/9912233].

\bibitem{Maeda:2000wr}
K.~i.~Maeda and D.~Wands,
Phys.\ Rev.\ D {\bf 62}, 124009 (2000)
[arXiv:hep-th/0008188].

\bibitem{Gen:2000nu}
U.~Gen and M.~Sasaki,
Prog.\ Theor.\ Phys.\  {\bf 105}, 591 (2001)
[arXiv:gr-qc/0011078].

\bibitem{Langlois:2000ns}
D.~Langlois, R.~Maartens and D.~Wands,
Phys.\ Lett.\ B {\bf 489}, 259 (2000)
[arXiv:hep-th/0006007];
\\
A.~V.~Frolov and L.~Kofman,
arXiv:hep-th/0209133.

\bibitem{Dubovsky:2000am}
S.~L.~Dubovsky, V.~A.~Rubakov and P.~G.~Tinyakov,
Phys.\ Rev.\ D {\bf 62}, 105011 (2000)
[arXiv:hep-th/0006046].

\bibitem{Langlois:2003dd}
D.~Langlois and M.~Sasaki,
arXiv:hep-th/0302069.

\bibitem{Langlois:2000iu}
D.~Langlois, R.~Maartens, M.~Sasaki and D.~Wands,
Phys.\ Rev.\ D {\bf 63}, 084009 (2001)
[arXiv:hep-th/0012044].

\bibitem{Gen:2002rb}
U.~Gen and M.~Sasaki,
Prog.\ Theor.\ Phys.\  {\bf 108}, 471 (2002)
[arXiv:gr-qc/0201031].

\bibitem{Bardeen:kt}
J.~M.~Bardeen,
Phys.\ Rev.\ D {\bf 22}, 1882 (1980).

\bibitem{Kodama:bj}
H.~Kodama and M.~Sasaki,
Prog.\ Theor.\ Phys.\ Suppl.\  {\bf 78}, 1 (1984).

\bibitem{Wands:2000dp}
D.~Wands, K.~A.~Malik, D.~H.~Lyth and A.~R.~Liddle,
Phys.\ Rev.\ D {\bf 62}, 043527 (2000)
[arXiv:astro-ph/0003278].

\bibitem{Sasaki:1986hm}
M.~Sasaki,
Prog.\ Theor.\ Phys.\  {\bf 76}, 1036 (1986).

\bibitem{KannoSoda}
S.~Kanno and J.~Soda,
Phys.\ Rev.\ D {\bf 66}, 043526 (2002)
[arXiv:hep-th/0205188];
\\
S.~Kanno and J.~Soda,
Phys.\ Rev.\ D {\bf 66}, 083506 (2002)
[arXiv:hep-th/0207029].

\bibitem{Kanno:03}
S.~Kanno and J.~Soda, hep-th/0303203.

\bibitem{Bateman:1969ku}
See e.g., H. Bateman, {\it Higher Transcendental Functions} Vol. I
(McGraw-Hill, New York, 1953).

\end{thebibliography}
\end{document}